\documentclass[12pt]{article}

\usepackage{latexsym}
\usepackage{amsmath}
\usepackage{epsf}
\usepackage{bm}
\usepackage{amsfonts}
\usepackage{amssymb}
\usepackage{graphicx,multicol}

\usepackage{color}
\definecolor{rosso}{cmyk}{0,1,1,0.4}
\definecolor{rossos}{cmyk}{0,1,1,0.55}
\definecolor{rossoc}{cmyk}{0,1,1,0.2}
\definecolor{blu}{cmyk}{1,1,0,0.3}
\definecolor{blus}{cmyk}{1,1,0,0.6}
\definecolor{bluc}{cmyk}{1,1,0,0.1}
\definecolor{verde}{cmyk}{0.92,0,0.59,0.25}
\definecolor{verdec}{cmyk}{0.92,0,0.59,0.15}
\definecolor{verdes}{cmyk}{0.92,0,0.59,0.7}

\def\circa#1{\,\raise.3ex\hbox{$#1$\kern-.75em\lower1ex\hbox{$\sim$}}\,}

\font\tenrsfs=rsfs10 at 12pt
\font\sevenrsfs=rsfs7
\font\fiversfs=rsfs5
\newfam\rsfsfam
\textfont\rsfsfam=\tenrsfs
\scriptfont\rsfsfam=\sevenrsfs
\scriptscriptfont\rsfsfam=\fiversfs
\def\mathscr#1{{\fam\rsfsfam\relax#1}}

\makeatletter

\newcommand{\be}{\begin{equation}}
\newcommand{\ee}{\end{equation}}
\newcommand{\beqy}{\begin{eqnarray}}
\newcommand{\eeqy}{\end{eqnarray}}
\newcommand{\p}{\partial}

\newcommand{\mx}{\mbox}
\newcommand{\mt}{\mathtt}

\newcommand{\bb}{\beta}
\newcommand{\ga}{\gamma}
\newcommand{\te}{\theta}

\newcommand{\de}{\delta}

\newcommand{\ze}{\zeta}

\newcommand{\e}{\epsilon}
\newcommand{\om}{\omega}
\newcommand{\Om}{\Omega}

\newcommand{\La}{\Lambda}

\newcommand{\n}{\nabla}

\newcommand{\ti}{\widetilde}
\newcommand{\bM}{\tilde{M}}
\newcommand{\tir}{\ti{r}}
\newcommand{\tik}{\ti{k}}

\newcommand{\rb}{\bar{r}}
\newcommand{\kx}{k_{\mt{max}}}
\newcommand{\bto}{\bar{\tau}_0}

\newcommand{\cE}{{\cal E}}
\newcommand{\cO}{{\cal O}}

\newcommand{\2}{\frac{1}{2}}
\newcommand{\3}{\frac{1}{3}}

\newcommand{\stwo}{\sqrt{2}}

\newcommand{\ra}{\rightarrow}
\newcommand{\Ra}{\Rightarrow}

\newcommand{\LF}{\left(}
\newcommand{\RF}{\right)}
\newcommand{\LT}{\left[}
\newcommand{\RT}{\right]}
\newcommand{\ie}{{\it i.e.\ }}

\newcommand{\vs}{\vspace{5mm}\\}

\begin{document}
\tolerance=100000
\thispagestyle{empty}
\setcounter{page}{0}

%\preprint{
IGPG-07/2-2
\vspace{1cm}

\begin{center}
{\LARGE \bf
``Swiss-Cheese'' Inhomogeneous 
 \\ [0.15cm]
 Cosmology \&
 \\ [0.18cm]
the Dark Energy Problem
}
\vskip 1cm
{
{\large Tirthabir Biswas\footnote{tbiswas@gravity.psu.edu}}$^{a,b}$,
 {\large Alessio Notari\footnote{notari@hep.physics.mcgill.ca}}$^{a}$,

\vskip 1mm
{\it $^a$ Physics Department, McGill University, \\ 3600 University Road, 
Montr\'eal, QC, H3A 2T8, Canada}
\vskip 3mm
{\it $^b$ Institute for  Gravitational Physics and Geometry\\
Department of Physics, PennState University\\
104, Davey Lab, University Park, PA 16802-6300
} 
\vskip 3mm
}
\vspace{1cm}
{\large\bf Abstract}
\end{center}
\begin{quote}
{\noindent

We study an exact swiss-cheese model of the Universe, where inhomogeneous LTB patches are embedded in a flat FLRW background, in order to see how observations of distant sources are affected. We find negligible integrated effect, suppressed by $(L/R_{H})^3$ (where $L$ is the size of one patch, and $R_{H}$ is the Hubble radius), both perturbatively and non-perturbatively. We disentangle this effect from the Doppler term (which is much larger and has been used recently~\cite{BMN} to try to fit the SN curve without dark energy) by making contact with cosmological perturbation theory.

}

\end{quote}

\newpage

\setcounter{page}{1}

%%%%%%%%%%%%%%%%%%%%%%%%%%%%%%%%%%%%%%%%%%%%%%%%%%%%%
\section{Introduction}
The fact that our universe seems to be accelerating presently has been one of the most remarkable and baffling  findings of recent cosmological observations. The origin of such a late time acceleration has been a source of intense research (see for example~\cite{taskforce}). The usual ways of explaining it are either to postulate the presence of a dark energy component (an unknown component with negative pressure), or to modify  gravity. The $\La$CDM model has become  the most popular choice being the most economical  model that is consistent with most cosmological observations, but it is safe to say that no completely satisfactory or compelling explanation has yet emerged. There are two aspects of the phenomenon which make it very hard to explain: the smallness problem, which essentially refers to the fact that the scale of new physics  required to explain dark energy,  $\sim  {\rm meV}$, is much smaller than expected. Secondly, if the equation of state for dark energy is so different from ordinary matter, as observations suggest, then one is left wondering about their approximate equality at the present epoch. In other words why did the dark energy start to dominate ``precisely'' today?  This is also known as the coincidence problem.

The problem of dark energy is so puzzling that it is perhaps worth taking a step back and investigate the assumptions based on which its existence has been inferred. The most direct evidence of dark energy comes from looking at distant supernovae, or more precisely at the relationship between the luminosity distance $D_L$ and the redshift $z$, upto $z\simeq \cO(1)$. It turns out that, in order to account for the observations, one requires a dominant dark energy component with an equation of state $\om\sim -1$ \cite{taskforce}. Indirect evidence for dark energy comes from CMB which concludes that our universe is flat, $\Om_{tot}\sim 1$, with $\Om_m\sim 0.3$ \cite{taskforce}. In other words,  non-relativistic matter (dark matter plus baryons) only accounts for around 30 \% of the total energy budget. The rest, 70\% of the energy content, can then be accounted for by   dark energy. The question is: are these conclusions robust or have  we made one assumption too many?  

For instance, an assumption in deducing the equation of state for dark energy from the supernova curve is that our universe is homogeneous and isotropic, and  that we can apply Friedmann-Lema\^itre-Robertson-Walker ({\it i.e.} homogeneous and isotropic) cosmology. Indeed, the FLRW model is a very good approximation in the Early Universe, as probed by the homogeneity of the CMB (density contrasts in  photons and matter are of the order  $10^{-5}$ and  $10^{-3}$ respectively). However, at low redshifts, the density contrast in matter grows up to values of ${\cal O}(1)$ and beyond: small scales become nonlinear first, and then more and more scales enter this regime. Today, scales of order of $60/h \, {\rm Mpc}$ (which is $2\%$ of the size of the horizon\footnote{In this paper we use $h=0.7$}) have an average density contrast of order 1. Structures observed with density contrast larger than $0.1$ (so, already in a mildly nonlinear regime) extend to few hundreds of Mpc (around $10\%$ of the size of the horizon). Since all our deductions about cosmology are based on light paths, and the presence of inhomogeneities  affects the background~\cite{buchert,rasanen,KMNR,notari}, the luminosity distance and the redshift, one can wonder whether  their effects can in fact account for the observed $D_L(z)$ curve, without resorting to dark energy.  One attractive feature of this possibility is that it has the obvious potential of addressing both the smallness and the coincidence problems. Hence, recently, such a possibility has received a considerable attention and there is an ongoing  debate as to whether the effect of inhomogeneities can be substantial. The purpose of this paper is to clarify this issue using a (semi-)realistic inhomogeneous cosmological model and performing both perturbative and non-perturbative analysis, both analytically (wherever possible) and numerically.

Before proceeding further, we should  point out that accounting for the luminosity distance-redshift curve is not enough, one also has to address CMB observations. This is a much more elaborate task which we leave eventually for future, but we only point out here that there are several degeneracies involved in CMB measurements which one may be able to exploit to fit the data with non-standard cosmological parameters (see for instance \cite{subir} for such an attempt).

Let us start by briefly going over the usual arguments as to why it is believed that the effect of  inhomogeneities is small and cannot account for  dark energy. Firstly,  it is generally argued that the Newtonian potential  $\Phi$, which characterizes the deviation from the  FLRW universe, is small even today because it is given approximately by the ratio $(L/R_H)^2\sim 10^{-5}$ at  the scale of nonlinearity, $L\sim 50/h\ {\rm Mpc}$. $R_H\sim 3000/h\ {\rm Mpc}$, denotes the Hubble radius. Since the corrections to redshift or luminosity distance is governed by $\Phi$, one expects them to be small. Secondly, even though light, as it passes through underdense or overdense regions is expected to get preferentially more red or blue shifted respectively, these contributions may cancel in the end. This intuition is corroborated  in first order perturbation theory where it is known that the corrections to the redshift, except terms which depend on position (linear Sachs-Wolfe effect) and velocity (Doppler term) of sources and observer, precisely vanish.  So how can inhomogeneities play any significant role?

To answer  this question it is crucial to realize that  the magnitude of corrections due to inhomogeneities on light paths basically depend on two dimensionless quantities, the length scale or period of fluctuations governed by the ratio $L/R_H$, and the amplitude of fluctuations, governed by $\de=(\rho-\langle\rho\rangle)/\langle\rho\rangle$, where $\rho$ is the energy density and $\langle...\rangle$ stands for a spatial average. Although the latter can be large, the former for realistic inhomogeneities is always small and therefore a perturbative analysis in the former is always good. It becomes important then to first identify the power of $L/R_H$ by which the corrections are suppressed. The second part is to determine  whether the ``prefactor'', which depends on $\de$ and can have non-perturbative corrections, could circumvent the suppression to give us something significant. These are the two issues we want to address in this paper. For this purpose it is convenient to  distinguish between two different ways that ``inhomogeneous cosmology'' may work:
\vs
{\it Scenario I, Local Effect:  There could be an effect due to our special position with respect to inhomogeneities.} For instance if we are located in an underdense region, then local observations (of redshift and luminosity distance) would be affected due to local variations in Hubble expansion. In particular what has been advocated in a long list of papers (see~\cite{celerier,Tomita00,Tomita01,wiltshire,moffat,alnes, reza, BMN}) is precisely such a model where the local underdensity can account for a larger (than average) local Hubble parameter thereby reconciling the discrepancy between nearby and distant supernovae, without dark energy. However, in order to make this work, one requires  the local underdense region to span at least upto the nearby supernovae \ie upto $z\sim 0.1$ which corresponds to $300-400\ {\rm Mpc}/h$. Although it has explicitly been demonstrated using exact Lema\^itre-Tolman-Bondi (LTB) solutions of Einstein's equations that such an inhomogeneity can in fact accommodate quite well the supernova curve~\cite{BMN}, the length scale required is too large as compared to the observed scale of non-linearity and therefore does not provide a completely satisfactory resolution\footnote{We can however point out the striking coincidence that the existence of structures of the same size could explain the large angle anomalies in the CMB sky~\cite{INOUESILK}, and that~\cite{hole} reports the observation of a 25\% lack of galaxies over a $(300/h{\rm Mpc})^3$ region.}. 

The reason, however, that  these models can avoid the arguments discussed above is two-fold. Firstly, as we clarify in our paper, the dominant local corrections go as $\n\Phi$, which is only suppressed as $L/R_H$ unlike $\Phi$. For scales as large as a few hundred Mpc, the suppression is therefore mild. Secondly, for local observations, since light only passes through underdense regions, there is no cancellation of effects and it is precisely the Doppler term (which is present already in perturbation theory, and which would average to zero for the entire universe) that contributes, enabling  one to account for the discrepancy between the nearby and distant supernovae \cite{BMN}. We also  show how using perturbative analysis one can reproduce the same results, so that there is nothing puzzling about this ``large'' local effect. Physically it amounts to saying that, if we live near the center of a large underdense region, all nearby sources are attracted towards its boundary: therefore they have a collective radial peculiar velocity which mimicks a large local value of $H$.
\vs
{\it Scenario II, Average Effect: light, while travelling though inhomogeneities, does not  see the ``average Hubble expansion'' but rather picks up systematic corrections to redshift and/or luminosity distance.} The hope is that such corrections, which are known to be present, could be non-negligible. This is a much more ambitious hope, since one wants to have a sizeable effect using only realistic nonlinear inhomogeneities (of order $50-60\ {\rm Mpc}/h$) to explain deviation of the $D_L(z)$ curve from the EdS (Einstein-de Sitter) universe. {\it A priori} there are at least three reasons that may make someone hopeful. Firstly, although it  is generally believed that $\Phi\ll 1$  in our universe, except very near black holes, to our knowledge there is no rigorous proof. Secondly, in the real universe, since voids occupy a much larger region as compared to structures, light preferentially travels much more through voids and this could negate the cancellation between voids and structures in quantities such as the redshift and the luminosity distance. Thirdly, since $\de\sim \cO(1)$ non-perturbative correction in $\de$ may enhance the corrections~\cite{notari}. We want to investigate whether any of these could give a sizeable effect.  

We choose to study  a swiss-cheese model containing tightly fitted spherical  patches\footnote{While completing the present paper, we noticed that a similar strategy has recently been followed in~\cite{greci}. However, we mainly focus on the calculation of the redshift, while~\cite{greci} focuses on the luminosity distance.}. Each of these patches contains a spherical void region and a thin dense spherical shell (much like the filamentary structure observed in the large scale distribution of galaxies) representing structures. Such a model has its short-comings: inside the patch the metric is spherically symmetric, which is not the case in our real universe, and spherical symmetry is known to produce surprising cancellations absent in real world\footnote{The very fact that the two metrics can be glued together is a manifestation of this special cancellation: {\it only} for spherical symmetry, due to Birkhoff's theorem, the metric outside the patch is insensitive to the details of what happens inside, but it is sensitive only to the total mass, exactly like Gauss' theorem in Newtonian physics. Another manifestation of this is, for example, the explosion of a supernova: it emits gravity waves only if the explosion is non spherical.}. Note that this feature suppresses any ``backreaction effect'' on the homogeneous regions: they still evolve {\it exactly} as a flat FLRW Universe. For this reason such  models do not incorporate any back-reaction effect, but only systematic corrections due to light propagation inside the patches (which is usually called Rees-Sciama effect). Secondly, we use an LTB metric which has a critical time when structures collapse to infinite density, unlike in the real universe, where the structures can virialize. 

Nevertheless, the model is semi-realistic and it is a start if one wants to estimate corrections to redshift and distances coming from inhomogeneities in our real universe. What we find is that although a local Doppler effect (\ie inside a single patch) is only suppressed by $L/R_H$, any average effect (\ie which survives even when one integrates over a whole patch) is much more suppressed and goes like $(L/R_H)^3$ confirming qualitative estimates present in literature~\cite{ReesSciama,MGSS}. There are then three key questions left to be answered. (i) Do these corrections add coherently which can perhaps increase the effect? The answer is yes, and in particular if one is interested in looking at corrections at $z\sim 1$, then effectively it goes like $(L/R_H)^2$ (but this is still too small to explain the SN curve). (ii) What about the $\de$ dependent prefactor, can it be large?  Our estimates indicate that the prefactor is $\cO(1)$, and it cannot circumvent the $(L/R_H)^3$ like suppression. (iii) What happens if one tries to generalize this analysis to perhaps non-spherical cases? Is it possible to get a lesser suppression? The answer to this question is well beyond the scope of this paper. We just point out that first order perturbation theory rules out an $\cO(L/R_H)$ correction, but if one could obtain a $(L/R_H)^2$ correction in a single patch, then the ``coherence effect'' along with an enhancement due to large density contrasts might lead to relevant effects. We leave further exploration of this possibility for future. 

In section~\ref{model}, we introduce our inhomogeneous swiss-cheese model of  universe and discuss qualitatively the expected corrections to redshift and distances. In section~\ref{smallusection} we discuss in detail the propagation of light, \ie redshift and luminosity distance, for profiles where curvature always remains small (although the density contrast can be $>\cO(1)$) along radial trajectories.  Next, in section~\ref{nonradial}, we discuss the connection between the LTB metric and perturbation theory about an FLRW Universe in order to estimate corrections in non-radial trajectories. In section~\ref{largeusection} we discuss profiles where curvature is large in some regions (and hence one may expect that $\Phi$ could be larger than what is usually advocated). Finally we conclude and we point out possible caveats in our model that  still leave hope of finding some non-trivial effect with more realistic inhomogeneities.

%%%%%%%%%%%%%%%%%%%%%%%%%%%%%%%%%%%%%%%%%%%
\section{The Swiss-Cheese Model of the Universe}
\label{model}
In this paper we will consider a swiss cheese type of model to describe our universe, where the inhomogeneities reside on spherical ``blobs'' cut out from a flat homogeneous Einstein-de Sitter (EdS) universe. We are going to use  LTB metrics to describe each of these blobs. As we will see later, LTB metrics can describe a wide range of density profiles (as a function of the radial coordinate), allowing us to mimick voids and structures. In principle one can consider these blobs to be randomly distributed, but since we are interested in looking at inhomogeneities  of a particular length scale, it is particularly useful to keep in mind a tight packing model, where the structures reside at the boundary of the blobs, while the insides  are mostly voids. This gives rise to a universe with continuous two dimensional structures where voids occupy the bulk of the space, somewhat similar to what observations suggest~\cite{voids}. We will however keep our analysis as general as possible with the aim of finding the maximal effect that can arise when  light  propagates through inhomogeneous blobs. 

We stress that there are two different effects: {\it Scenario I.} This is present when the observer lives {\it inside}  one of these patches and corresponds to a large local correction to the Hubble diagram (see~\cite{BMN} for an analysis of how observations can look like in this case), due to peculiar velocities. To investigate these kinds of effects it is suitable to place the observer at the center of a patch. {\it Scenario II.} The second effect is a net correction that a photon experiences when passing through a blob: in order to capture this,  we put  observers and sources {\it outside} the blobs, in the FLRW region, or equivalently at two diametrically opposite points on the boundary of the patch.

To have a picture of our swiss-cheese model it is important to realize that one can consistently choose a coordinate system where the boundaries of each of these blobs are constant in comoving coordinates. No matter (dust) flows in or out of the blobs, and with time the only thing that happens is that the physical radius of these blobs expands, exactly obeying the EdS expansion which arise out of the junction conditions. This fact enables us to    consistently paste these blobs onto the homogeneous FLRW metric. To understand this procedure better let us start with a brief review of LTB metrics. In the subsequent subsection we will discuss the junction conditions.

%%%%%%%%%%%%%%%%%%%%%%%%%%%%%%%
%\setcounter{equation}{0}
\subsection{LTB Metrics, a Short Review}

The LTB metric (in units $c=1$) can be written as
\begin{equation}
ds^2=-dt^2 + S^2(r,t)dr^2 + R^2(r,t)(d\theta^2 + \sin^2 \theta 
d\varphi^2) , \label{eq:14}
\,  \end{equation}
where we employ comoving coordinates ($r,\theta,\varphi$) and
proper time $t$. 

Einstein's equations with the stress-energy tensor of 
dust, imply the following constraints:
\begin{eqnarray}
S^2(r,t) &=& {R^{'2}(r,t)\over {1+2E(r)}} , \label{eq:15}   \\
{1\over 2} \dot{R}^2(r,t) &-& {GM(r)\over R(r,t)}=E(r) , \label{eq:16} 
\\
4\pi \rho (r,t) &=& {M'(r) \over R'(r,t) R^2(r,t)}  \label{eq:17}
\, , 
\end{eqnarray}
where a dot denotes partial differentiation with respect to $t$ and a prime 
with respect to $r$. The quantity $\rho (r,t)$ is the energy density of the matter, 
and $G\equiv 1/m_{Pl}^2$ is Newton's constant. 
The functions $E(r)$ and $M(r)$ are left arbitrary. The function $E(r)$ is  related to the spatial curvature, while 
$M(r)$ basically corresponds to the mass inside a sphere of comoving 
radial coordinate $r$ \cite{LTB,celerier}. If $M(r)$ is an increasing function, it can always be chosen to be
\be
 M(r)=\frac{4 \pi}{3}M_0^4 r^3 \, ,
\ee
by a suitable reparametrization of the radial coordinate and we will adopt this convention for the rest of the paper. Here $M_0$ is an arbitrary mass scale.

One can easily verify  that, depending upon the sign of $E(r)$, (\ref{eq:16}) admits solutions analogous to the flat, closed and open universes~\cite{LTB}\footnote{The reader should not be confused by the terminology 
which is usual in the literature: for instance a model with $E(r)>0$ can easily mimick a flat FLRW model. It is 
also possible in principle to have a mixed model with some regions where 
$E(r)<0$ and other regions where $E(r)>0$.}, but for our purpose it will be sufficient to only look at the open case, $E(r)>0$:
\begin{eqnarray}
R&=&{GM(r)\over 2E(r)} (\cosh u-1) , \label{eq:18} \\
t-t_b(r)&=&{GM(r)\over [2E(r)]^{3/2}}(\sinh u-u) \, , 
\end{eqnarray}
where $t_b(r)$, often known as the ``bang-time'', is another arbitrary 
function of $r$. It is 
interpreted, for 
cosmological purposes, as a Big-Bang singularity surface  at which $R(r,t)=0$. This is analogous to the scale-factor 
vanishing at the big bang singularity in the homogeneous FLRW models. It is easy to see that the function $t_b(r)$ becomes negligible at late 
times, $t\gg t_b(r)$, and therefore we can just ignore it and set $t_b(r)=0$ for all $r$. 

For later convenience we enumerate  the  simplified equations for our model:
\be
R(r,t)={2\pi r\over 3 k(r)}(\cosh u-1)\, ,
\label{Ru}
\ee
\be
\tau^3\equiv \bM t={\pi\stwo \over 3 k(r)^{3/2}}(\sinh u-u)\label{tu} \, ,
\ee
where we have introduced the conformal time $\tau$, the ``curvature function''
\be
k(r)\equiv {E(r)\over \bM^2 r^2}\,
\, , \ee 
and
\be
\bM\equiv {M_0^2\over m_{Pl}} \, .
\ee
%%%%%%%%%%%%%%%%%%%%%%%%%%%%
\subsection{Junction Conditions and Constraints on Curvature}    \label{swiss}
Before we proceed further it is useful to choose conventions 
which simplify both analytic and numerical calculations. We  
choose $M_0$ in such a way that the ``coordinate density'' $M_0^4$ 
coincides with the average (EdS) density $\rho_0$, at the present time $t_0$:
\be
M_0^4=\rho_0={ M_p^2\over 6 \pi t_0^2}
\label{convention}
\, . \ee
Or, in other words
\be
\tau_0^3=t_0\bM = {1\over \sqrt{6\pi}} 
\label{tau0}
\, . \ee
Essentially such a convention implies that the coordinate $r$ is approximately the physical distance between the comoving coordinate point and the center, today.
With this convention we also find that the present Hubble scale $H_0$ (and hence the Hubble radius $R_H$)  of the average model is simply related to the parameter $\bM$:
\be
R_H^{-1}= H_0\equiv {2\over 3t_0}=\sqrt{\frac{8\pi}{3} } \bM 
\, . \ee

Next, let us look at the curvature function. There is a procedure on how to match the blobs to a homogeneous metric consistently~\cite{reza2}, according to which the curvature function has to satisfy 
\be
k'(L)=0 \, .
\ee 
We denote by $L$ the  comoving radius of the boundary where the blob meets the homogeneous background. In general a given blob maps to a particular FLRW universe. Note that FLRW universes are characterized by a single parameter: it could be the value of  the ``curvature abundance'' $\Om_k$ at some given epoch (labeled by the value of the Hubble parameter, for instance). One can check that if we use the above normalization (\ref{convention}) then 
\be
k(L)={4\pi\over 3}\Om_k\, , \qquad \mx{ for } \, |\Om_k|\ll 1
\, , \ee
where $\Om_k$ is the ``curvature abundance'' of the homogeneous universe at the same epoch.
In particular, if we want the curvature to vanish, in order to be consistent with CMB, then we would  need $\Om_k\ll 1$, and for simplicity we will only look at profiles where it exactly vanishes. Thus our profile has the following property
\be
k(L)=k'(L)=0
\label{boundary}
\, . \ee

There is also an additional constraint on curvature that comes from requiring continuity at the origin~\cite{Flanagan}:
\be
k'(0)=0
\label{center}
\, . \ee
It is now clear that as long as we have profiles, or $k(r)$'s,  which obey (\ref{boundary}) and (\ref{center}), we can embed these spherical inhomogeneous blobs inside EdS consistently. In particular we can study how light travels through a series of these blobs.
%%%%%%%%%%%%%%%%%%%%%%%%%%%%%%%%%%%%
\subsection{Qualitative Discussion of Corrections}

Having chosen the model, we want to study observational quantities of distant sources: the redshift $z$ and the luminosity distance $D_L$. Our aim will be to check whether there is any non-perturbative large correction to the FLRW predictions, or if we essentially recover results from perturbation theory.
Here we discuss what are the possible outcomes.

In order to understand the corrections to various quantities due to inhomogeneities it is important to realize that there are two important parameters in the problem: the curvature inside a patch (which can be parameterized for example by the maximal value $k_{\mt{max}}$ of the function $k(r)$), and the ratio $L/R_H\sim \bM L$. While the former can  most directly be related to the amplitude of density fluctuations (as we will see, the density contrast goes like $\de\propto k_{\mt{max}}\tau^2$, for small $\de$), the latter governs the period or length scale of fluctuations.  During non-linear structure formation $\de\sim \cO(1)$, but a realistic value of the nonlinear length scales gives at most $(\bM L)\sim 0.01$. One can push this limit to somewhat larger values but still it remains a small parameter. Therefore it plays the most important role in determining the magnitude of correction that one expects from inhomogeneities. In particular it is of crucial importance to determine the power of $\bM L$ that appears in the correction to quantities such as $\de z$ (the correction to redshift w.r.t. the FLRW result) or $\de D$ (the correction to the angular or luminosity distance w.r.t. the FLRW result). The next important question is to see whether the $\kx$ dependent factor in the corrections can ever overcome the suppression due to $(\bM L)$. 

For clarity, we will approach the problem in  stages. We will start as a first step by considering radial photon trajectories passing through profiles where the curvature always remains ``small''(but $\delta$ can be large). As we will see, ``small curvature'' implies that the Newtonian potential $\Phi\sim \delta (L/ R_H)^2$, and therefore is small as it is generally believed. Also, radial trajectories necessarily have to  pass through both voids and structures, which may have cancelling effects on the corrections.  So, we are possibly underestimating the corrections, but in this special case one can at least find  analytical solutions for  $t(r)$ (time along the photon trajectory), $z(r)$ and $D_L(r)$ (and hence also $D_L(z)$),  and therefore precisely determine how  the corrections to these quantities depend on $\kx$ and $\bM L$. We find that there are local corrections (Doppler effect) which go like $L/ R_H$ and therefore can be significant. In fact  these are the kind of effects which are exploited in \cite{BMN} to explain away dark energy with an $L\sim 300-400 {\rm Mpc}/h$, \ie {\it scenario I}. However, overall corrections that are relevant for {\it scenario II} are much more suppressed and go like $(L/ R_H)^3$, which is too small to be significant for supernova cosmology.  The $\kx$ dependent prefactor cannot overcome this suppression. This is also born out by numerical solutions of the equations of motion.  

Next we check whether there can be any enhancement of the effect by considering  non-radial trajectories which do not have to pass through ``equal amounts'' of void and structure. Numerical solutions show that one does not find any significant difference in the magnitude of correction, and this can be further corroborated using  perturbation theory. Thus, at least for low curvature profiles, we are tempted to conclude that the average corrections are suppressed by  $(L/ R_H)^3$, and there is no significant non-perturbative enhancement in $\kx$ which can remotely overcome such a suppression.

Next, we consider profiles which have some regions with high curvature. Unfortunately in this regime one does not have complete  analytic control. It is not clear, for example,  whether it is possible to go to a Newtonian gauge where $\Phi$ remains small everywhere, but this is perhaps also the reason to be hopeful. It turns out that it is still possible to estimate the corrections that one can expect in such generic situations. Unfortunately, it seems that the corrections look very similar to the small curvature case. In other words the Doppler contribution goes as $L/ R_H$, but the overall correction is again suppressed as $(L/ R_H)^3$. The only difference, it seems, is that the non-perturbative effects in $\kx$  can give rise to a larger pre-factor.  However, this enhancement still falls far short of what could be significant for dark energy. 

Finally, we look at a series of patches to check whether the correction in each patch adds coherently or not. This turns out to be the case and thus if we are interested in estimating corrections for a total distance that correspond to  redshifts of $\cO(1)$, which is  approximately the same as the horizon scale, the average correction gets a $R_H/L$ enhancement. This leads ultimately to an overall effect  $\sim (L/R_H)^2$, but still clearly too small. 

In the rest of the paper we provide details of the above discussion.
%%%%%%%%%%%%%%%%%%%%%%%%%%%%%%%%%%%%

\setcounter{equation}{0}
\section{Small Curvature Regime} \label{smallusection}
To get an intuitive picture of the swiss-cheese LTB model we will first employ what we call ``small curvature'' approximation~\cite{BMN}, which allows us to study everything analytically.  The strength of the approximation lies in the fact that it can accurately describe the dynamics even when $\de \rho/\rho\gg 1$ \cite{BMN}, and thereby allows us to study the effect of non-linear structure formation on the different physical quantities. We will see in the next section that this is also a regime when perturbation theory is definitely valid, \ie one can compute the Newtonian potential and show that it is small.  In this section we will focus on only radial trajectories, where a cancellation of effects between structures and voids may be at work. So, we are perhaps underestimating the effects of inhomogeneities but the hope is that we will have an analytical understanding of how the corrections depend on the two parameters, $\bM L$ and $\kx$ and whether non-perturbative effects in the latter can overcome the suppression due to the former.
%%%%%%%%%%%%%%%%%%%%%%%%%%%%%%%%%%%%%%%%%%
\subsection{Small-$u$ Approximation and Density Fluctuations}\label{structure}
In  ~(\ref{Ru}) and~(\ref{tu}), if we only keep next to leading terms in $u$ we find \cite{BMN}
\be
R(r,t)\approx {\pi\over 3}\ga^2\tau^{2} r\left[1+R_2\ga^2\tau^2 k(r)\right]
\label{R}
\, , \ee
where $\ga$ and $R_2$ are constants:
\be
R_2\equiv {1\over 20}\qquad \mx{ and } \qquad \ga\equiv \LF{9\sqrt{2} \over \pi}\RF^{1/3}
\label{gamma}
\, . \ee
One can check  that this approximation is valid (\ie $u$ 
is small) when 
\be
u\approx u_0\equiv \ga\tau\sqrt{k(r)}\ll 1
\label{v}
\, . \ee

We will first consider density profiles where this condition is satisfied for all $r$ inside the patches. Now, to compute $\rho(r,t)$ we have to compute $R'$. From (\ref{R}) we find
\be
R'(r,t)={\pi\over 3}\ga^2\tau^{2}\left[1+R_2\ga^2\tau^2\left(rk(r)\right)'\right]  \label{Rprimo}
\, . \ee
From (\ref{eq:17}) we have
\be
\rho(r,t)={M_0^4\over 6\pi(\bM t)^2\left[1+ R_2\ga^2\tau^2k(r)\right]^2\left[1+R_2\ga^2\tau^2\left(rk(r)\right)'\right]}
\, . \ee
We note that the correction term in the second factor of the 
denominator is actually proportional to $u_0^2$
$$
R_2\ga^2 k(r)\tau^2=R_2u_0^2\ll1 \, ,
$$
and therefore it can be ignored consistently within our ``small-$u$'' 
approximation.
Thus we have
\be
\rho(r,t)={M_0^4\over 6\pi(\bM t)^2\left[1+R_2\ga^2\tau^2A(r)\right]}\,\, ,\mx{ where 
}A(r)\equiv\left(rk\right)'
\label{matterdensity}
\, . \ee

We observe that the FLRW behaviour for 
the density is given by the prefactor multiplying the denominator, while the fluctuations are provided by the presence of 
$A(r)$ in the denominator. It is clear that the density contrast ($\de$) defined in the usual 
manner
\be
\de\equiv {\rho(r,t)-\langle \rho \rangle(t)\over \langle \rho 
\rangle(t)} 
\, , \ee
is controlled by
\be
\e(r,t)\equiv R_2\ga^2\tau^2A(r)
\, . \ee
One obtains a simple relation
\be
\de= -{\e\over 1+\e}=-{R_2\ga^2\tau^2A(r)\over 1+R_2\ga^2\tau^2A(r)}
\label{contrast}
\, . \ee
When $A(r)$ is close to its maximum value we have a void, while when it is 
close to  its minimum, it signals an overdensity.
Note that~\cite{BMN}, when $\epsilon\ll 1$, $\de\sim \e$ and we are in the linear regime: an initial density 
fluctuation grows at small times as 
$\tau^2\sim t^{\frac{2}{3}}$, in agreement with the prediction of cosmological 
perturbation theory.
When $\epsilon$ is no longer small the density contrast grows rapidly (and this result is  the same as found within 
the Zeldovich approximation \cite{zeldovich}).
%%%%%%%%%%%%%%%%%%%%%%%%%%%%%%%%%%%
\subsection{Radial Photon Trajectories} \label{photon}
As we will see, we are also able to analytically study the trajectory of a light ray, which propagates radially inside one patch.

The first step is to solve  the null geodesic equations. In 
other words, we have to obtain $t(r)$ for a photon trajectory 
converging at say $r=r_0$ at time $t=t_O$ 
(from now on, the subscript $O$ will denote the observer).
The equation for this radial photon trajectory is simply given by
\be
{dt(r)\over dr}=\pm{R'(r,t(r))\over \sqrt{1+2E(r)}} \, . \label{tphoton}
\ee 
The $\pm$ depends on the direction of photon propagation. Now we are only going to be interested in placing the observer either at the boundary, $r=L$ (and to be definite say at $\te=\phi=0$) when investigating average effects ({\it scenario II}), or at the center, $r=0$, while looking at local effects ({\it scenario I}).  It is then convenient to define a different variable $\ti{r}\in [-L,L]$, such that $\ti{r}=r$ along the interval $r\in [0,L],\ \phi=\te=0$, while $\ti{r}=-r$ for the portion of the path $r\in [0,L],\  \phi=\pi,\ \te=0$. Defining $k(-\tir)=k(\tir)$ for negative arguments\footnote{Since the derivative of $k(r)$ vanishes at $r=0$, $k(\tir)$ remains a smooth function.}, one then finds that the differential equation that is valid through out the path for photons which are travelling along increasing values of $\tir$ is given by
\be
{dt\over d\tir}={R'(\tir,t(\tir))\over \sqrt{1+2E(\tir)}} \, .
\label{t-radial}
\ee
The prime now indicates a differentiation with respect to $\tir$. In the rest of the paper we are going to drop the tilde from $r$, but the readers should note that when we talk about negative values of $r$, we are really talking about $\tir$.

Now, we are interested in solving the above equation in the small curvature regime, \ie keeping only upto linear terms in $k(r)$. One can, in fact, find an analytical solution (see appendix~\ref{A-smallu}) which looks like
\be
\tau=\tau_F(r)+\de\tau(r)
\, , \ee
where $\tau_F$ is the FLRW trajectory
\be
\tau_F(\rb)=\tau_0+{\pi\over 9}\ga^2(\rb-\rb_O)\equiv \bto+{\pi\over 9}\ga^2\rb
\, , \ee
and 
\be
\de\tau={\pi\over 9}\ga^2\LF R_2\ga^2\bto^2\LT\rb k+2{\pi \ga^2 \over 9\bto}k\rb^2 +\LF{\pi \ga^2 \over 9\bto}\RF^2k\rb^3-2{\pi \ga^2 \over 9\bto}\int d\rb\ k\rb\RT-{6\over 5}\int d\rb\ k\rb^2\RF
\label{deltau}
\, , \ee
is the correction coming from inhomogeneities. Here $\tau_0$ is an integration constant denoting the conformal time at $r=r_0$ and we have defined the dimensionless variable
\be
\rb=\bM r
\, , \ee
for convenience. 

It is now instructive to look at all the different terms and estimate them. The first three terms all vanish when the source is at the boundary $r=-L$ although, inside the patch, these terms are non-vanishing and  dominate over the contributions coming from the integrals.  The contribution of these first three terms can be combined conveniently to give
\be
\tau(\rb)\approx \tau_F(\rb)\LT 1-{\pi R_2\over9} \ga^4 \rb\tau_F(\rb) k(\rb)\RT
\label{tau-local}
\, . \ee
As we will see later, this corresponds to the Doppler term in perturbation theory and in principle can give rise to large correction to the redshift. In fact it is this contribution which has been exploited in the ``void models'' to explain away the dark energy problem. However,  this effect only lasts as long as one is inside the patch and therefore to account for dark energy  one has to have a really large void, $L\sim 300 \, {\rm Mpc}/h$, which would at least include the nearby supernovae. We leave for future to study how likely such an explanation can be \cite{future}. 

It is easy to see what is the  suppression in powers of $L/R_H$ of the terms in~(\ref{deltau}): it sufficient to count the powers of $r$.
 The Doppler term~(\ref{tau-local}) is therefore proportional to ($\bM L$), but since it originates as a total derivative it vanishes outside the patch; also in the language of  first order perturbation theory a Doppler term  cannot contribute to any average effect. Thus although we have verified this for radial trajectories in a spherically symmetric model (and for only a single patch), the robustness of perturbative arguments suggests that there are no average corrections proportional to ($\bM L$) in general.

Next, let us look at possible contributions of the order $(\bM L)^2$ which one can naively suspect to arise from the first integral in (\ref{deltau}). However, in the special case of radial trajectories at least, this vanishes when integrating from boundary to boundary  because the integrand is an odd function of $\rb$ ($k(r)$ is even by definition). However, one may hope that in a more general case (relaxing the spherical symmetry or considering non-radial trajectories) such a term might give a non-vanishing contribution. We will come back to this point later in section \ref{nonradial}. 

Thus in the end we find that the only non-zero contribution comes from the last term and is proportional to  $(\bM L)^3$. One can in fact come up with an upper bound for such a contribution (see appendix \ref{A-smallu}) in our swiss-cheese model:
\be
{|\de \tau|\over \tau_0}\leq{4\pi\over 45} \sqrt{6\pi}(\bM L)^3\approx 1.2 (\bM L)^3 \, .
\label{smallubound}
\ee

%%%%%%%%%%%%%%%%%%%%%%%%%%%%%%%
\subsection{Redshift and Luminosity Distance}
In the previous subsection we have found $t(r)$ along a photon 
trajectory. In this subsection we obtain the redshift $z(r)$ corresponding to a source located at $r$. The differential equation governing this 
relation is given by~\cite{celerier}:
\be
{dz\over dr}= - {(1+z)\dot{R}'\over \sqrt{1+2E}}
\label{redshift}
\, . \ee

In order to solve (\ref{redshift}) we need to compute $\dot{R}'$. In the small-$u$ approximation we have
\be
\dot{R}'={2\pi\ga^2\bM\over 9\tau}[1+2 R_2 \gamma^2 \tau^2 (r k)'] \label{Rpdot}
\, . \ee
First let us focus on computing ``local effects'' (which show up when the observer is located inside the patch, {\it e.g.} in the center). Inside the patch the largest correction in~(\ref{Rpdot}) is the term linear in $\rb$ and hence we can ignore $E\sim \rb^2$ in~(\ref{redshift}) (since the patch is assumed to be much smaller than the horizon). In this approximation, using (\ref{tau-local}), one can solve for redshift analytically (see appendix \ref{App-redshift}): 
\be
1+z=\LT{\tau_0\over \tau_F(\rb)}\RT^2\exp\LT {- 4\pi R_2\over9} \ga^4 \rb\tau_F(\rb) k(\rb)\RT \, 
\label{red-inside}
\, . \ee 

Let us try to understand the corrections (\ref{red-inside}). We note that, for an open universe $k>0$ and $\rb<0$ for our trajectory, so that the correction is always positive. In other words one always perceives a greater redshift as compared to FLRW. Why is that? In an open LTB patch one has to have an underdense region at the center and an overdense region near the boundary. Thus let us consider the simplest profile (the argument goes through for more complicated profiles as well), which is to have an underdense region at the center, followed by an overdense region which matches to the average density region at the boundary. In this case, the quantity $rk(r)$ in the exponent generically reaches a maximum somewhere in the middle and then falls to zero at the boundary. The maximal deviation from FLRW is therefore attained at this maximal value which coincides with the average density ($[rk(r)]'=0$ corresponds to having no corrections in the density function, see (\ref{matterdensity})). Physically: starting from the center till the average density point the space describes a void which is being stretched more than the homogeneous FLRW. Hence we get an additional redshift. After that point the space describes a structure which is contracting, which keeps compensating for the extra redshift and finally at the boundary precisely cancels it.

To see why  the ``local effect'' in redshift is significant, let us look at small redshifts ({\it i.e.} small $\bar{r}$). In this case we find
\be
z\approx {2\pi\ga^2\rb\over 9\tau_0}(1+2R_2\ga^2 k \tau_0^2)\equiv z_F(r)(1+2R_2\ga^2 k \tau_0^2)=z_F(r)(1+2R_2u_0^2) \, ,
\ee
where $z_F(r)$ is the FLRW value of the redshift. Just as an estimate, taking $u_0^2\sim 1$, we therefore find a 10\% correction. If one allows for profiles where curvature can be larger, \ie $u_0^2>1$, one typically gets an even larger correction. In any case it is already clear that locally one can get significant effects from inhomogeneities.

However the situation is very different when one tries to compute the ``net'' effect of the inhomogeneities on redshift as seen by an observer outside the patch, which is relevant for the success of {\it scenario II}. We notice that, at the boundary $r=-L$, the redshift in~(\ref{red-inside}) coincides with the FLRW result since the curvature term vanishes at the boundary. So these terms do not lead to any net or average effect, as was also found in the case of $\de \tau$ in the previous subsection. This can again be traced back to the fact that the corrections linear in $\rb$ come from total derivative Doppler terms.

To find net effects we have to go to higher order terms (in $\rb$).  The equation for redshift (\ref{redshift}) in the small-$u$ approximation gives us
\be
{dz\over 1+z}= {2\pi\over 9}{\ga^2\over\tau}\left[1+2R_2 \ga^2\tau^2 (\rb k)'-k\rb^2\right]d\rb
\, . \ee
This, up to linear terms in $k$, reads
\be
\ln (1+z)={2\pi\ga^2\over 9}\LT \int {d\rb\over \tau_F+\de\tau}+2R_2 \ga^2\int d\rb\ \tau_F (\rb k)'-\int d\rb\ {k\rb^2\over \tau_F}\RT
\label{z-full}
\, . \ee
We note here that for the purpose of estimating net effects we have to place the observer and the source at two diametrically opposite points in the boundary, so that now the integral runs from $-L$ to $L$. This essentially guarantees that the correction can only go as $(\bM L)^3$. The linear terms are absent because they originate as total derivatives and $k$ vanishes at the boundary, while the quadratic terms vanish because the integrand is odd in $\rb$. This is obvious for the last two terms in (\ref{z-full}). The first term may look more complicated because  we find expressions containing double integrals (using~\ref{deltau})). However, it is easy to see that all these double integrals can be converted to single integrals by suitable integration by parts. As a result a typical correction term looks like
$$
\rb^n|_E\int_{-L}^{L} d\rb\ \rb^mk(\rb)
$$
where the subscript ``$E$'' stands for quantities evaluated at the source (emitter) of the photons, and $m,n$ are integers, with $m>0$. Again when one is integrating from boundary to boundary the integral vanishes if $m$ is odd. Thus the largest correction arises when $n=0$ and $m=2$, giving rise to a correction proportional to $(\bM L)^3$. 

Actually, one can check that even the $(\bM L)^3$ term vanishes at linear order in $k(r)$, consistent with first order perturbation theory (see next section), but for higher orders in $k(r)$ this does not happen. Thus the above analysis is instructive to see why we will in general get a $(\bM L)^3$ correction in the average effects, while we get a $(\bM L)$ correction for local effects. Similar arguments can also  be applied to estimate corrections in $\de D$ as well, yielding  similar results. Thus, to conclude, we expect local effects in the $D_L(z)$ curve  to be suppressed only linearly in $(\bM L)$, while average corrections receive a large cubic suppression.

%%%%%%%%%%%%%%%%%%%%%%%%%%%%%%%%%%
\section{Non-Radial Trajectories \& Perturbation Theory}\label{nonradial}
%%%%%%%%%%%%%%%%%%%%%%%%%%%%%%%%%%%%%%%%
The radial trajectories that we discussed in the previous section necessarily go through structures as well as voids and this, one may suspect, could provide an additional mechanism for cancellation of effects. We now want to test whether considering more general trajectories  leads to a larger effect in $\de z$ or $\de D_A$, or not. This is relevant for our universe because voids today occupy a much larger volume as compared to structures and therefore one does not expect the photons to see ``equal amounts of over and underdense regions''\footnote{Since in open LTB models we can only have structures near the periphery, we can only have non-radial trajectories which preferentially cross more of structures and not the other way round. If we consider a closed LTB model we could engineer the other case. However, it should not matter, since the purpose is really to estimate the correction.}.  

In this subsection we consider non-radial trajectories which enter the boundary not directed along the radius, but at an angle to it. We then keep track of the redshift and luminosity distance as the photon exits through another point at the boundary. Since the spherical inhomogeneous patches have an overdensity near the periphery, by varying the entry angle we can vary the relative amounts that the photon spends in  over and underdense regions and thereby break any cancellation effects between over and underdense regions. In particular there are trajectories where the photon only sees an overdensity and therefore we expect to see an enhancement in the corrections. The crucial question is: by {\it how much?} Unfortunately, we cannot solve the equations analytically as we did in the radial case, but we have to resort to numerics. Numerical solutions show no relevant variation between corrections in radial versus non-radial case (We show the comparisons in figures~(\ref{1patch}) and (\ref{1patchAng})). Thus we conclude that the effects must still be suppressed by an $(\bM L)^3$ factor as before.

In fact, this result can be confirmed  using perturbative estimates. As we will see, small curvature also implies that the Newtonian potential is small and therefore one can hope perturbation theory to be a useful guide. Moreover, these estimates do not rely on trajectories being  radial and is equally applicable for non-radial ones. Therefore below we clarify the correspondence between standard perturbation theory and LTB models in small-$u$ approximation and also estimate corrections that one can generally (radially and non-radially) expect in redshift and  distances. 
%%%%%%%%%%%%%%%%%%%%%%%%%%%%%%%%%%%
\subsection{Newtonian Potential}
Our aim in this section is to try to cast the metric~(\ref{eq:14}) in the small-$u$ limit in a perturbative form:
\be
ds^2=a^2(\tau)[-(1+2\Phi)d\tau^2+(1-2\Phi)dx^2] 
\label{newtonian}
\, , \ee
and verify that $\Phi\ll 1$ and  that it is  time independent. We will see that this is  at least possible in the small-$u$ and small-$E$ limit. Rather than trying to find the appropriate gauge transformations (which is explained in the appendix~\ref{gauges}), our strategy would be to compute for the LTB metric the two gauge-invariant combinations that are well known in first order perturbation theory. If these two potentials turn out to be equal, time independent and small, then it will be clear that indeed the LTB metric can be cast into~(\ref{newtonian}).

In the small-$u$ limit we find that the LTB metric is given by
$$
ds^2={9\tau^4\over \bM^2}\left\{-d\tau^2+\LF{\pi\ga^2\over 3}\RF^2{\left[1+R_2\ga^2\tau^2\left(rk(r)\right)'\right]^2\over 1+2k\rb^2}d\rb^2+\LF{\pi\ga^2\over 3}\RF^2\rb^2 \left[1+R_2\ga^2\tau^2 k(r)\right]^2d\Om^2\right\}
$$
Further, assuming $E\ll1$ we have
\be
ds^2={9\tau^4\over \bM^2}\left\{-d\tau^2+\left[1+2R_2\ga^2\tau^2\left(\tir \tik(\tir)\right)'-20 R_2\ga^2\tir^2\tik(\tir)\right]d\tir^2+\tir^2 \left[1+2R_2\ga^2\tau^2 \tik(\tir)\right]d\Om^2\right\}
\label{metric-k}
\, . \ee
Here we have defined the new dimensionless coordinate:
\be
\ti{r}={\pi\ga^2\over 9} r\bM\Ra d\ti{r}={\pi\ga^2\over 9} \bM dr
\, . \ee
Also, the prime now refers to differentiation with respect to $\tir$. One can think of (\ref{metric-k}) as a truncation of the LTB metric keeping only upto linear terms in $k(r)$. 
Our aim now is to find the gauge invariant variables for the above metric.

Now, in general, any metric containing only scalar fluctuations can be written as 
\be
ds^2=a^2(\tau)\{-(1+2\psi) d\tau^2+2\p_i\omega d\tau dx^i+[(1-2\phi)\de_{ij}+D_{ij}\chi]dx^idx^j\}
\, , \ee
where we have defined the operator
\be
D_{ij}\equiv \p_i\p_j-\3\de_{ij}\n^2
\, . \ee
Transforming from the Cartesian to the radial coordinate system we have generically
\be
ds^2=a^2(\tau)\left[-(1+2\psi)d\tau^2+2\omega'(r)drd\tau +\LF1-2\phi+{2\over 3}\cE\RF dr^2+\LF 1-2\phi-\3{\cal E}\RF r^2d\Om^2\RT
\, , \ee
where we have defined
\be
\cE\equiv \chi''-{\chi'\over r}  \label{Echi}
\, . \ee

It is now easy to read off the different potentials corresponding to the LTB metric:
\be
\phi=-\3R_2\ga^2\LF3\tau^2k+\tau^2\tir k'-10 R_2\ga^2\tir^2 k\RF \label{phiLTB}
\, , \ee
and
\be
{\cal E}=2R_2\ga^2\tau^2 \tir k'-20 R_2\ga^2\tir^2 k   \label{curlyE}
\, , \ee
while the potentials $\psi,\ \omega$ vanish, and the background scale factor is given by
\be
a(\tau)={3\tau^2\over \bM}
\, . \ee

What is observationally important are of course the gauge invariant combinations (that we denote by $\Phi$ and $\Psi$), which in linear order look like
\be
\Psi=\psi+{1\over a}\LT\LF-\omega+{\chi_{\tau}\over 2}\RF a\RT_{\tau}={1\over 2a}(\chi_{\tau}a)_{\tau}=\2\LF \chi_{\tau\tau}+\chi_{\tau}{a_{\tau}\over a}\RF
\, , \ee
and 
\be
\Phi=\phi+{1\over 6}\n^2\chi-{a_{\tau}\over a}\LF \omega-{\chi_{\tau}\over 2}\RF =\phi+{1\over 6}\LF \chi''+{2\chi'\over r}\RF+{\chi_{\tau}\over 2}{a_{\tau}\over a}
\, . \ee
$\Phi$ and $\Psi$ coincide with the potentials $\psi$ and $\phi$ in the Newtonian gauge (\ref{newtonian}) and for consistency with first order perturbation theory they should be equal~\cite{MFB}. It is a straightforward exercise to compute these quantities using~(\ref{Echi}),(\ref{phiLTB}) and~(\ref{curlyE}). Indeed they turn out to be equal and one finds
\be \label{gravpot}
-\Psi=-\Phi= 6R_2\ga^2\int d\tir\ \tir k(\tir)<  {3\over 5} k_{\mt{max}} (\bM L)^2 
\, , \ee
which is indeed small when the length scales of inhomogeneities that we are considering are small compared to the Hubble radius.

Therefore, one can estimate the Rees-Sciama effect coming from perturbation theory and check that it is in agreement with our findings, as we will do in the next subsection.

%%%%%%%%%%%%%%%%%%%%%%%%%%%%%%%%%%%%%%%%%%%%%%%%%5
\subsection{Comparison with Perturbative Rees-Sciama Calculation}

It is well-known that there is an overall effect on the redshift of a
photon passing through a structure (usually called Rees-Sciama effect~\cite{ReesSciama,MGSS}).
This has been estimated previously by many authors using different
techniques (qualitative arguments~\cite{ReesSciama}, 
Void models~\cite{vishniac}, swiss-cheese models~\cite{MGSS} , LTB models~(\cite{panek,fullana,arnau}),second order calculations~\cite{MM}).
The aim of this section is to review the logic behind this calculation, and compare the second order result with our findings.

First we present a qualitative order of magnitude estimate, based on the smallness of the perturbation of the metric.

Using a weak-field method (in which one assumes the metric~(\ref{eq:14}), and the gravitational potential $\Phi$ to be small) it can be shown that corrections to the redshift of a photon go like:
\begin{equation}
\frac{\delta z}{(1+z)} \simeq \Phi_E-\Phi_O +v^i_E e_i-v^i_O e_i +\int
d\tau
\frac{\partial \Phi}{\partial \tau} \, ,    \label{RS}
\end{equation}
where $v_{O,E}$ is the peculiar velocity respectively of the observer and of the emitter in the Newtonian frame.
The first two terms are the so-called Sachs-Wolfe effect (difference of gravitational potential of emitter and observer).
The velocity terms are due to the peculiar motion of emitter and observer: they represent the Doppler effect.
The third term is an integrated term, which is present only if the first order gravitational potential evolves with time (Integrated Sachs-Wolfe effect).
Now, it is well-known that in linear perturbation theory (on scales 
where
we can treat $\delta$ linearly) $\Phi$ is time independent, during
matter domination. So, the integrated effect has to be zero, and the 
only
effect is due to the position and the motion of the observer (and of the source) in the Newtonian coordinates.
This means that, even if a photon passes through an inhomogeneous 
blob,
this does not matter for its final redshift, since the only nonzero 
effect
comes from the starting and the arrival points and not from the path
between. 

Thus  perturbation theory tells us that at first order there are no average effects, however there can be local effects! In fact, the Doppler terms in (\ref{RS}) give a contribution  which goes as the radial velocity of the emitter, which in turn goes as $\sim \n\Phi$. This gives rise to precisely the kind of terms that we found in our previous section while discussing corrections to $\de\tau$ and $\de z$ along radial trajectories in the small $u$ approximation. 

Next, let us look at how ``net corrections'' can nevertheless arise  at the nonlinear level in $\delta$.
This can be estimated as follows~\cite{panek}:
\begin{equation}
\left. \frac{\delta z}{(1+z)} \right|_{overall} \simeq
\int d\tau \frac{\partial \Phi}{\partial \tau}
\simeq \frac{\Phi}{tc}{\Delta t}
\, , \end{equation}
where $t_c$ is the typical time scale of change of the gravitational
potential and $\Delta t$ is the travelling time inside a structure.
Now,  $\Delta t$ is roughly equal to the size of the structure $L$ and
$t_c$ can be estimated as $t_c=\frac{L}{v_c}$, where $v_c$ is the 
typical
velocity inside the structure, leading to
\begin{equation}
\frac{\delta z}{(1+z)}\approx \Phi \, v_c
\, .\end{equation}
If the structure is still not virialized a good estimate for $v_c$ is
$v_c=|\nabla \Phi|$.
Instead, if the structure is virialized a good estimate is given by
$v_c^2\simeq \Phi$.
In both the cases, we may express the result in terms of the density contrast $\delta\equiv\frac{\rho-\rho_0}{\rho_0}$. Using the Poisson equation
$\nabla^2\Phi=4\pi G \, \bar{\rho} \, \delta \, a^2$
we get an estimate in terms of $\delta$ and $L$.

For non-virialized structures:
\begin{equation}
\frac{\delta z}{(1+z)}\approx \delta^2 \left(\frac{L}{R_H}\right)^3 \, . \label{nonvir}
\end{equation}
For a virialized structure:
\begin{equation}
\frac{\delta z}{(1+z)}\approx \delta^{3/2} \left(\frac{L}{R_H}\right)^3 \, .
\end{equation}

We can check now, using a systematic expansion of the metric up to second order, if we recover the result~(\ref{nonvir}). We follow the treatment of~\cite{MM} to compute the redshift of a photon in a Universe with only dust, perturbed to second order. 
The four-velocity of the dust is expanded as:
\begin{equation}
U^\mu=\frac{1}{a}\left(\delta^\mu_0 + v^{(1)\mu}+\frac{1}{2}
 v^{(2)\mu}+\ldots \right).
\end{equation}
This is subject to the normalization condition $U^\mu
U_\mu=-1$.

The perturbed spatially flat conformal metric has the metric elements: 
\begin{equation}\label{eq:m1}
g_{00}=-\left(1+2 \psi^{(1)}+ \psi^{(2)}+\ldots ,\right)\;,
\end{equation}
\begin{equation}\label{eq:m2}
g_{0i}=\omega_i^{(1)}+\frac{1}{2} \omega_i^{(2)}+\ldots ,
\end{equation}
\begin{equation}\label{eq:m3}
g_{ij}=\left(1-2 \phi^{(1)} -\phi^{(2)}\right)\delta_{ij}+
\chi^{(1)}_{ij}+\frac{1}{2}\chi^{(2)}_{ij}+\ldots ,
\end{equation}
where\footnote{Indices are raised and lowered
using $\delta^{ij}$ and $\delta_{ij}$, respectively.}
$\chi^{(r)i}_{i}=0$ and the functions $\psi^{(r)}$, $\omega^{(r)}_i$,
$\phi^{(r)}$, and $\chi^{(r)}_{ij}$ represent the
 $r$-th order perturbation of the metric.

The first order calculation gives the well-known terms:
\begin{equation}
\delta z^{(1)}=\psi^{(1)}_{E}-\psi^{(1)}_{O}+
v^{(1)i}_{O} e_i-v^{(1)i}_{{E}} e_i-I_1(\lambda_{E}).\label{dt1p} \, ,
\end{equation}
where $e^i$ is the unit vector in the direction of observation and  
\begin{eqnarray}
I_1(\tau_{E})&=&-\int_{\tau_{O}}^{\tau_{E}}
d\tau \left[\frac{1}{2}\frac{\partial}{\partial \tau} A^{(1)} \right]
\, , \end{eqnarray}
and $A^{(1)}\equiv \psi^{(1)}+\phi^{(1)}+\omega^{(1)}_i e^i-
\frac{1}{2}\chi^{(1)}_{ij}e^i e^j$.

This corresponds to~(\ref{RS}), but it is a more general expression, valid in {\it all} gauges. As we just saw, the integrated effect is zero.

Then, we may study the second order expression for $\delta z$. 
The general expression is very complicated and contains a lot of terms.
However, we are interested in studying the overall effect from boundary to boundary of a single patch.
At the boundary all gravitational potentials and all velocities are zero (and even second derivatives of the gravitational potential).
So the result reduces to
\begin{eqnarray}
\delta z^{(2)}|_{{\rm boundary \, to \, boundary}}=- I_2(\tau_{E})&=&-\int_{\tau_{O}}^{\tau_{E}}
d\tau \left[\frac{1}{2}\frac{\partial}{\partial \tau} A^{(2)} \right]
\, , \end{eqnarray}
where $A^{(2)}\equiv \psi^{(2)}+\phi^{(2)}+\omega^{(2)}_i e^i-
\frac{1}{2}\chi^{(2)}_{ij}e^i e^j$.

Due to the spherical symmetry of our configuration the tensor traceless part of the metric, $\chi^{ij(2)}$, vanishes.

Then at second order, in the so-called Poisson gauge (a generalization of the Newtonian gauge to second order~\cite{MM}) one has:
\begin{eqnarray}
 \psi^{(2)}&=&\phi^{(2)}=\tau^2 \left(\frac{1}{6}\varphi^{,i}\varphi_{,i}
-\frac{10}{21}\Upsilon_0\right) \, , \nonumber \\
\nabla^2 \omega^{(2)i}&=&-\frac{8}{3}\tau \left(\varphi^{,i}
\nabla^2\varphi-\varphi^{,ij}\varphi_{,j}+2\Upsilon_0^{,i}\right),
\end{eqnarray}
where
\begin{equation}
\nabla^2 \Upsilon_0=-\frac{1}{2}\left((\nabla^2\varphi)^2-
\varphi_{,ik}\varphi^{,ik}\right).
\end{equation}
However one may notice, by simple power counting, that the vector term contains one power less of spatial derivatives, which means that it is suppressed with respect to the scalar contribution. 
So, the dominant contribution is:
\begin{equation}
\delta z^{(2)}_{scalar}=- 2 \int \, d\tau  \, \tau \left(\frac{1}{6}\varphi^{,i}\varphi_{,i}-\frac{10}{21} \Upsilon_0 \right) \, ,
\end{equation}

Let's now try to work out explicitly $\delta z^{(2)}$, using our spherically symmetric potential $\phi(r)$, given by~\ref{gravpot}.
%Using the following identities
%\begin{eqnarray}
%\nabla^2 f(r)&=&\frac{1}{r^2}\frac{\partial}{\partial r}\left( r^2 f(r) \right) \nonumber \\
%\partial_j (f(r))&=& \frac{x_j}{r} f'(r)
%\end{eqnarray}
We get:
\begin{equation}
\delta z^{(2)}_{scalar}=- 2 \int \, d\tau  \, \tau \left(\frac{1}{6} \varphi'^2-\frac{10}{21} \Upsilon_0 \right) \, , \label{RS2}
\end{equation}
where
\begin{equation}
\Upsilon_0(r)=-\int^r \frac{d \bar{r}}{\bar{r}^2} 
\int^{\bar{r}} d\tilde{r} \left[(\varphi'(\tilde{r}))^2+2 \tilde{r} \varphi''( \tilde{r})\varphi'( \tilde{r}) \right]
\, .\end{equation}

%Then the vector term gives instead:
%\begin{equation}
%\delta z^{(2)}_{vector}=\frac{8}{3} \int \, d\tau \, \left[\frac{1}{r^2} \int^r d\tilde{r}\Upsilon_0(\tilde{r})+\int^r \frac{d\bar{r}}{\bar{r}^2}\int^{\bar{r}} d\tilde{r} \, \tilde{r} \, \varphi'^2(r) \right]
%\end{equation}

Using~(\ref{gravpot}) one finds
\begin{equation}
\delta z^{(2)}_{scalar}=- 2 \int \, d\tau  \, \tau \left( 6 R_2^2 \gamma^4 (k r)^2-\frac{10}{21} \Upsilon_0 \right) \, ,
\label{deltaz2}
\end{equation}
where
\begin{equation}
\Upsilon_0(r)=- 36 R_2^2 \gamma^4 \int^r \frac{d \bar{r}}{\bar{r}^2}
\int^{\bar{r}} d\tilde{r} \left[3 (k \tilde{r})^2+2 \tilde{r}^3 k k'\right]
\, .\end{equation}
We notice that the largest non-zero term in (\ref{deltaz2}) goes like $k^2(\bM L)^3$ corroborating the qualitative estimates for non-virialized structures (\ref{nonvir}). We also notice that the cubic dependence in $\bM L$ agrees with our in the previous section in the small-$u$ limit\footnote{The $k^2$ dependence originates because the first non-zero correction in redshift is obtained from second order perturbation theory. The correction in the proper time, $\de \tau$ is however linear in $k$ as derived in the small-$u$ limit, and one can also check this perturbatively.}.

%%%%%%%%%%%%%%%%%%%%%%%%%%%%%%%%%%
\section{Large Curvature Regime}  \label{largeusection}
%%%%%%%%%%%%%%%%%%%%%%%%%%%%%%%%%%%%%%
In the previous sections we have found that, at least in the small curvature regime, there is no significant ``average'' correction to the luminosity-redshift relation. However, in general, LTB profiles can have some regions with large curvature. Thus it is important to understand at least how the corrections scale with $L$ in the general case. Also, here it is not clear anymore that one can go to a Newtonian gauge where $\Phi\ll1$ everywhere and so, as suggested in the introduction, perhaps we have the greatest chance of uncovering a large correction in such set-ups. Also, in this general case we can no longer restrict ourselves to linear terms in $\kx$ and therefore any potentially large non-perturbative correction in $\kx$ could show up here.

To understand some of the generic features of the model let us fall back to the defining equations (\ref{Ru}) and (\ref{tu}). It is easy to see that (\ref{v}) can be inverted to obtain $u$ as a power-series in $t$:
\be
u=u_0\LF1 + c_2 u_0^2 + c_4 u_0^4 + \dots\RF
\, , \ee
where $u_0$ is given in terms of $t$ via (\ref{v}). We had computed the coefficient $c_2=-1/60$ to obtain the small-$u$ limit. It is instructive to see how this function behaves for very large $u$:
in this regime we can approximate
$$\sinh u,\cosh u\ra \2e^u.$$
From (\ref{tu}) we then get
\be
\lim_{u_0\ra \infty}u\longrightarrow 3\ln u_0-\ln 3 \, .
\label{largeu} \ee 
Thus, initially $u$ is proportional to $u_0$, but later it increases only logarithmically.

\begin{figure}
\begin{center}
\includegraphics[width=0.6\textwidth]{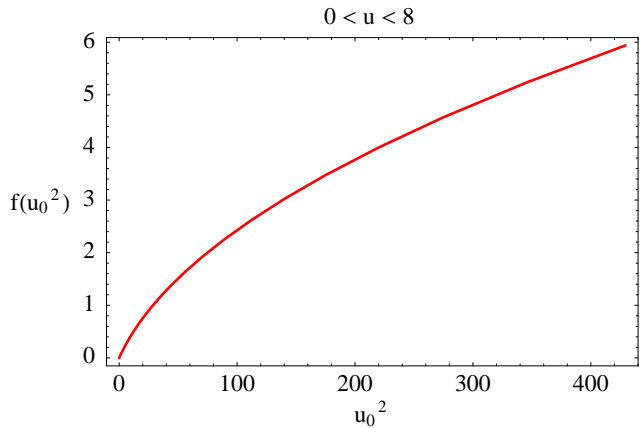}

\caption{ \label{effefig}
\small
Here we plot $f(u_0^2)$, obtained combining~(\ref{Ru}),(\ref{tu}),(\ref{v}) and (\ref{R-u0}).
}
\end{center}
\end{figure}

One can also obtain a general expression of $R(r,t)$ as a power series:
\be
R(r,t)=\3\pi \ga^2r\tau^2\LF 1+R_2u_0^2+R_4u_0^4+\dots\RF\equiv \3\pi \ga^2r\tau^2\LF 1+f(u_0^2)\RF
\label{R-u0}
\, . \ee
We show the behaviour of the function $f(u_0)$ numerically in fig.~\ref{effefig}. The FLRW geometry corresponds to just including the first term, and the correction, $f$, can be large if $u_0$ is large (since $f$ is a monotonic function, as it can be seen in fig.~\ref{effefig}) . Therefore, at first sight, one might also  expect reasonably large corrections. Going to the  large $u$ limit, where one can compute $R$ approximately, such a reasoning indeed seems plausible. By substituting (\ref{largeu}) in (\ref{Ru}) we see that 
\be
R\longrightarrow {\pi r\over 3k}e^u=\3\pi \ga^2r\tau ^2\LF{u_0\over 3}\RF
\qquad \Ra \qquad  f=1-{u_0\over 3} 
\, , \ee

%%%%%%%%%%%%%%5

suggesting an $\cO(1)$ departure from the FLRW universe. In particular it also becomes unclear whether there is a suitable gauge transformation which can take such a metric to the Newtonian gauge with a small Newtonian potential. Thus it becomes very important to check whether indeed one obtains large deviations to $D_L(z)$ relation.
%%%%%%%%%%%%%%%%%%%%%%%%%%%%%%%%%%%%%
\subsection{Estimating Corrections}
What we want to do here is to estimate the magnitude of corrections that one can expect for generic profiles. In particular we want to find corrections to the photon trajectory governed by equations (\ref{tphoton}) and~(\ref{redshift}). In the previous analysis what we did was to keep the lowest order terms in both $k(r)$ and $\rb$ which  gave us non-zero effects. Here we will consider the full power series expansion in the ``amplitude'' $k(r)$ (since curvature can now become large) but still keep the lowest non-trivial power in $\rb$. Now, for the purpose of estimation, it turns out that we can ignore  the denominator (the $1/\sqrt{1+E}$ term) in (\ref{tphoton}) (see appendix \ref{A-largeu} for details). The evolution equation then reads 
\be
{dt\over dr}\approx R'\approx\frac{\pi}{3} \ga^2\tau^2\LF 1+\sum_{n} R_{2n}\ga^{2n}\tau^{2n}(rk^n)'\RF \label{evoluz}
\, . \ee

\begin{figure}
\begin{center}
\includegraphics[width=0.6\textwidth]{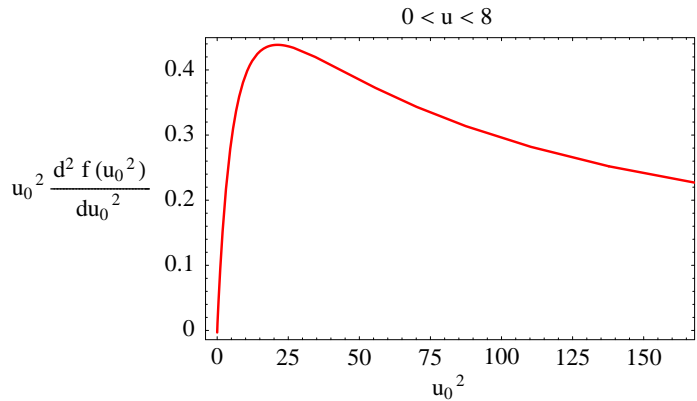}
\caption{ \label{df2} \small
Here we plot $u_0^2 \frac{d^2f(u_0^2)}{du_0^2}$, obtained combining~(\ref{Ru}),(\ref{tu}),(\ref{v}) and (\ref{R-u0}).
}
\end{center}
\end{figure}

This equation can be solved in much the same way as it was done in the small-$u$ case. The results are also almost identical, the linear terms in $r$ vanish because they originate from total derivative terms, and the quadratic terms vanish because the integrands become odd. (Instead of terms linear in $k(r)$ we have to evaluate integrals which involve some powers of $k(r)$, but this clearly does not change any of the above arguments.) Thus the largest contribution is again proportional to $(\bM L)^3$. By solving~(\ref{evoluz}) one finds the following correction to $\tau_F$ (see Appendix~\ref{A-largeu})
\be
{\de \tau\over\tau_0}\approx -\LF{\pi \ga^2 \over 9\tau_0}\RF^3\sum_1^{\infty} R_{2n}\ga^{2n}\tau_0^{2n}2n(2n-1)\int_{-L}^L d\rb\ \rb^2k^n
\, . \ee
One can also obtain an upper bound for the above correction (see appendix \ref{A-largeu}):
\be
{|\de \tau|\over \tau_0}\leq (\bM L)^3\LF{\pi \ga^2 \over 9\tau_0}\RF^2\LF u_0^2{d^2f\over du_0^2}\right)_{\mt{maximum}}
\, . 
\ee
At this point it is important to realize that the function $f(u_0^2)$ is a well defined function which does not depend on the specific profiles (see figure \ref{effefig}) and in fact the function  $u_0^2{d^2f\over du_0^2}$ has a maximum $\sim 0.5$ (see figure~(\ref{df2})). Using this we get the following upper bound for $\de\tau$  
\be
{|\de \tau|\over \tau_0}\leq 1.5 (\bM L)^3
\, , \ee
not much different from what was obtained in the small-$u$ limit (\ref{smallubound}). We chose to study $\delta\tau$ (even if it not directly observable) to make the discussion simple, but we expect similar corrections to redshift and luminosity distance. Numerical solution of the equation of motion bear this out and this is what  we discuss now.

%%%%%%%%%%%%%%%%%%%%%%%%%%%%%%%%%%
\subsection{Numerical Results}
%%%%%%%%%%%%%%%%%%%%%%%%%%%%%%%

In this subsection we discuss the numerical solutions for Einstein equations and for a light ray in such a background.

First, we have calculated if there is any sizeable effect with a ``large curvature'' profile in a single patch.
We have chosen the following profile:
\begin{equation}
k(r)=k_0 \left[ \left(\frac{r}{L}\right)^2-1 \right]^2 \, ,   \label{numprofile}
\end{equation}
which satisfies conditions~(\ref{boundary}) and~(\ref{center}), and where the amplitude $k_0$ is chosen in such a way that we have a desired density contrast at the present time $t_0$.
We have then glued this single patch to a flat FLRW background.

\begin{figure}
\includegraphics[width=0.7\textwidth]{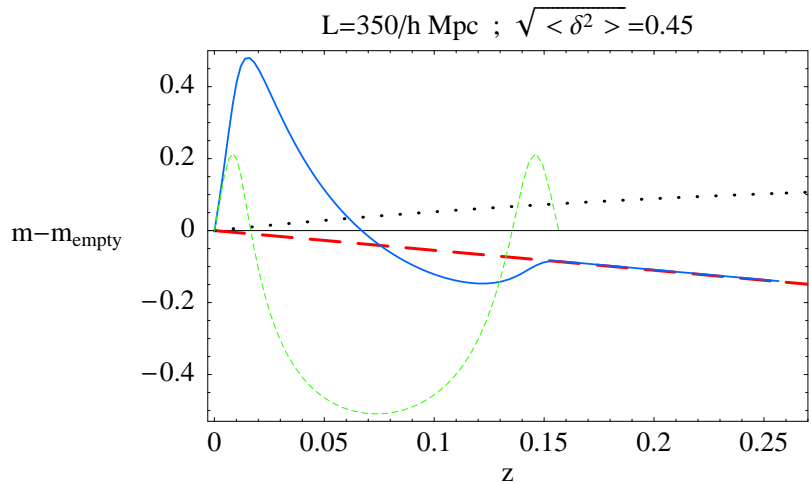} 
\caption{  \label{1patch}
{\small Here we plot $\Delta m$ (the difference in magnitudes, $m\equiv 5 Log_{10}D_L$, between a given model and a reference empty open universe) vs.~$z$ (redshift).
The red long-dashed line is an EdS Universe (only dust), the black dotted line is the $\Lambda CDM$ ``concordance model'' and the blue solid line is a model with light going {\it radially} through one inhomogeneous patch (the observer sitting outside the patch).
There is no visible correction to the EdS model at $z\gtrsim 0.15$ (objects behind the patch), since the correction is very small. We also show the density contrast (on the same scale) that the photons see along their trajectory (green thin short-dashed line).}}
\end{figure}

\begin{figure}
\includegraphics[width=0.7\textwidth]{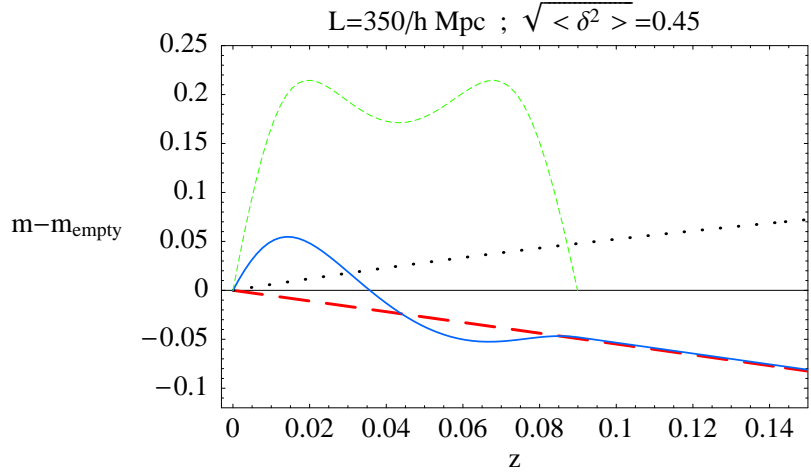}
\caption{ \small \label{1patchAng}
Here we plot $\Delta m$ (the difference in magnitudes between a given model and a reference empty open universe) vs.~$z$ (redshift).
The red long-dashed line is an EdS Universe (only dust), the black dotted line is the $\Lambda CDM$ ``concordance model'' and the blue solid line is a model with light going {\it non-radially} through one inhomogeneous patch (the observer sitting outside the patch).
There is no visible correction to the EdS model at $z\gtrsim 0.15$ (objects behind the patch), since the correction is very small. We also show the density contrast (on the same scale) that the photons see along their trajectory (green thin short-dashed line).}
\end{figure}

In fig.~(\ref{1patch}) we have run the code which calculates the redshift  and the luminosity distance of the photon for a radial trajectory in this profile, while in fig.~(\ref{1patchAng}) we have considered a photon propagating non-radially in the same background. The deviation due to the Doppler effect is visible, but no net effect is well visible in the plots.

\begin{figure}
\includegraphics[width=0.7\textwidth]{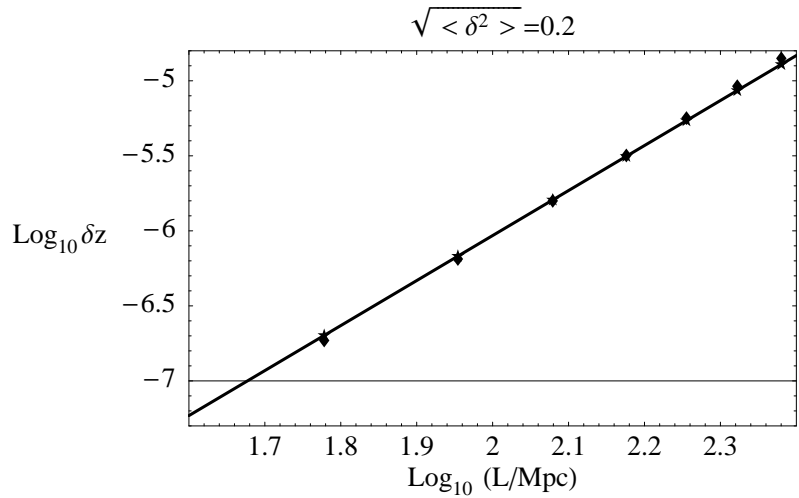}
\caption{ \label{elle} \small
In this plot we show the net correction to the redshift ($\delta z$) as a function of the size of the patch ($L$), for an observer sitting outside the patch.
The triangles are the numerical results, while the solid line shows a cubic dependence ($\delta z\propto L^3$). As expected the points follow a cubic dependence. It is interesting to note that, for a patch with $\delta\approx 0.2$ and a radius of about $200/h$ Mpc, there would be a ${\cal O}(10^{-5})$ correction for the redshift along the line of sight, which would consequently be visible as a secondary effect in the CMB (see also~\cite{INOUESILK}).}

\end{figure}

Nonetheless the overall net effect is non-zero and we have considered its dependence on $L/r_{hor}$, by running the program with the profile~(\ref{numprofile}) for several values of $L$, keeping $k_0$ fixed. We have plotted the net correction in redshift ($\delta z$), as a function of $L$ in fig.~(\ref{elle}) which verifies the cubic dependence derived analytically.

\begin{figure}
\includegraphics[width=0.7\textwidth]{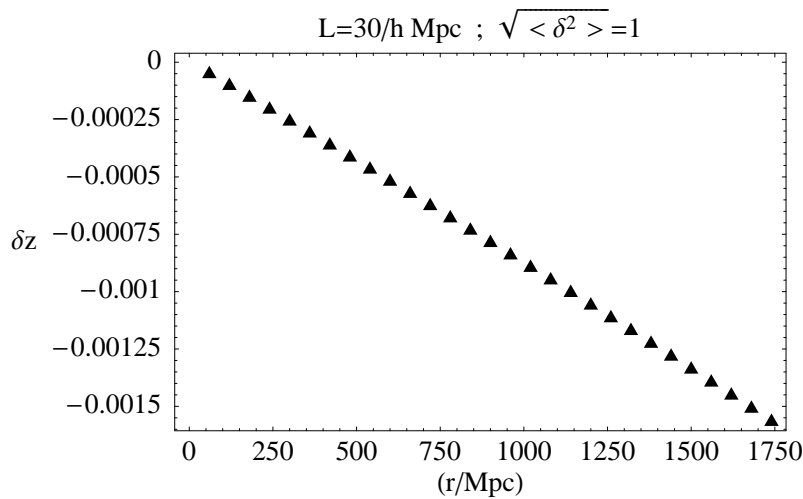}
\caption{ \label{manypatches} \small
In this plot we show the net correction to the redshift ($\delta z$) for a photon going through multiple adjacent inhomogeneous patches. All the patches are identical (with physical radius $L=30$ Mpc and density contrast $\delta\approx 1$).
Each point in the plot corresponds to the correction in redshift with respect to the homogeneous case ($\delta z\equiv z-z_{FLRW}$), when the photon comes out from one patch and enters the next one.}
\end{figure}

Finally we have considered a ``tight-packing'' model with several identical patches, and a photon propagating radially through all of them. We plot the result in fig.~(\ref{manypatches}). As one can see the correction is cumulative.


%%%%%%%%%%%%%%%%%%%%%%%%%%%%%%%%%%%%%
\section{Conclusions}
\label{conclusions}
We have studied photon propagation in a swiss-cheese model in which spherical patches are embedded in a flat FLRW Universe, both when the curvature of the patches is small and  large.
In both cases the effects are very small if the observer sits outside the patch~(integrated effect), while they are large if he sits inside (local effect).

We have checked this in the perturbative regime (small curvature), analytically and numerically and we have also made contact with perturbation theory around an FLRW background. In this way we have identified the contribution of local terms (peculiar position and velocity of source and observer) and integrated effects. As  is well-known, integrated effects appear only at second order, and this is in agreement with the fact that we find very small overall effects if the observer sits outside the patch. The nontrivial check that we have performed is to study the same thing in a large curvature regime. Both by analytical estimates and numerical solutions we do not find significant enhancements due to non-perturbatively large curvature terms. We also obtained the dependence  of the small net effect on the size of the patch, both in the perturbative (small curvature) and non-perturbative (high curvature) regimes, and we find a cubic suppression in $L/R_{hor}$.

There might still be two logical possibilities to obtain a larger average effect, that we have not explored in this paper. The first is to include somehow the possibility that structures virialize: the reason one may expect a difference as compared to collapsing structures is because the structures no longer contract after virialization, while the void regions keep on expanding. One may therefore expect a photon passing through both structures and voids to experience a comparatively larger redshift.  
The second possibility would be to break spherical symmetry: because of it, in this paper, we could not include backreaction effects of the local patches on the ``global'' FLRW evolution, and therefore this leaves open the possibility that  significant backreaction effect could appear in a more general case.

Finally we have clarified the origin of large local corrections in LTB (when the observer sits inside a patch), as a Doppler effect due to the peculiar motion of nearby sources. This effect has been used~(\cite{BMN} and references therein) to argue for the possibility of a larger than average local Hubble expansion and thereby explain the supernova data without dark energy, as a mismatch between local observations of the Hubble parameter and distant supernovae (outside the local patch).
This idea works, but it requires the existence of a very large underdense patch (the minimal radius of the patch has to extend at least up to $z\lesssim 0.1$, which corresponds to $300-400/h $ Mpc, with an average density contrast of roughly $\delta= - 0.3$ ) and it requires that we sit near the center\footnote{We postpone to future work a study of how close to the center we need to be, to be consistent with CMB observations.}. While this is in contrast with standard structure formation scenario, we can however point out the striking coincidence that the existence of structures of the same size has been invoked by~\cite{INOUESILK}, in order to explain the large angle anomalies in the CMB sky. It is also intriguing the fact that~\cite{hole} reports the observation of a 25\% lack of galaxies over a $(300/h {\rm Mpc})^3$ region, which may signal the existence of such structures.

\section{Acknowledgements}

We thank Reza Mansouri for useful discussions.

%%%%%%%%%%%%%%%%%%%%%%%%%%%
% Appendices
%%%%%%%%%%%%%%%%%%%%%%%%%%%
\appendix
\setcounter{equation}{0}
\section{``Average'' Corrections to $\tau$ in Small-$u$ Approximation}\label{A-smallu}
In the small-$u$ approximation one has (\ref{R})-(\ref{Rprimo})
$$
R={\pi\over 3}\ga^2r\tau^2(1+R_2 u_0^2)={\pi\over 3}\ga^2r\tau^2(1+R_2 \ga^2\tau^2 k)
$$
$$
R'={\pi\over 3}\ga^2\tau^2[1+R_2 \ga^2\tau^2 (rk)']
$$
and 
$$
\dot{R}'={2\pi\over 9}\ga^2{\bM\over\tau}\left[1+2R_2 \ga^2\tau^2 (rk)'\right]
$$

The evolution equation for $\tau$ is then given by (\ref{t-radial})
\be
{d\tau\over dr}={{\pi\over 9}\ga^2\bM[1+R_2 \ga^2\tau^2 (rk)']\over \sqrt{1+2k(\bM r)^2}}\approx {\pi\over 9}\ga^2\bM[1+R_2 \ga^2 \tau^2 (rk)'-k(\bM r)^2]
\, , \ee
where we have assumed that the photons are travelling from negative towards positive values of $r$. The above equation  leads us to the iterative expression
\be
 \tau=\tau_0+{\pi\over 9}\ga^2(\rb-\rb_O)+{\pi\over 9}\ga^2\LT R_2 \ga^2\int_O^S d\rb\ \tau^2(\rb k)'-\int_O^S d\rb\ k\rb^2\RT\equiv \tau_F+\de\tau
\, , \ee 
where $\tau_0$ is an integration constant denoting the conformal time at $r=r_0$ and $\de \tau$ denotes the corrections to the FLRW trajectory coming from the integrals. Now, since we are only interested upto terms linear in $k$, we can replace $\tau$ by $\tau_F$ inside the integrals, so that we have the approximate result
\be
\de\tau=\tau-\tau_F={\pi\over 9}\ga^2\LT R_2 \ga^2\int d\rb\ \tau_F^2(\rb k)'-\int d\rb\ k\rb^2\RT\equiv {\pi\over 9}\ga^2(I_1+I_2)
\, . \ee
where we have defined
$$
I_2\equiv -\int d\rb\ k\rb^2 \, ,
$$
and
\begin{eqnarray}
I_1&=&R_2\ga^2\bto^2\int d\rb\ \LT1+{\pi \ga^2 \rb\over 9\bto}\RT^2(\rb k)'=R_2\ga^2\bto^2\int d\rb\ \LT1+2{\pi \ga^2 \rb\over 9\bto}+\LF{\pi \ga^2 \rb\over 9\bto}\RF^2\RT(\rb k)' \nonumber \\
&=&R_2\ga^2\bto^2\LT\rb k+2{\pi \ga^2 \over 9\bto}k\rb^2 +\LF{\pi \ga^2 \over 9\bto}\RF^2k\rb^3-2{\pi \ga^2 \over 9\bto}\int d\rb\ k\rb-2\LF{\pi \ga^2 \over 9\bto}\RF^2\int d\rb\ k\rb^2\RT \, . \nonumber
\end{eqnarray}
In the last step we have used integration by parts and assumed that the observer is located either at the center where $r=0$, or at the boundary, $r=L$ where $k=0$. Combining $I_1$ with $I_2$ we have
\be
\de\tau={\pi\over 9}\ga^2\LF R_2\ga^2\bto^2\LT\rb k+2{\pi \ga^2 \over 9\bto}k\rb^2 +\LF{\pi \ga^2 \over 9\bto}\RF^2k\rb^3-2{\pi \ga^2 \over 9\bto}\int d\rb\ k\rb\RT-{6\over 5}\int d\rb\ k\rb^2\RF
\, . \ee
The first three terms correspond to local effects which vanish at the boundary. The fourth term (integral) vanishes when evaluated from $-L$ to $L$. So the net  correction comes from the last term and is indeed very small. In fact one can easily find an upper bound as follows:
\be
|\de \tau|={2\pi\over 15}\ga^2\int d\rb\ \rb^2k\leq{4\pi\over 45}\ga^2\kx (\bM L)^3
\, . \ee
attained by $k(r)$ inside the profile.  
Now, for small-$u$ we have the obvious bound
\be
u_0^2=\ga^2\tau_0^2 k(r)\leq 1\qquad \Ra \qquad \ga^2 \kx\leq {1\over \tau_0^2} 
\, . \ee
 Thus we get an upper bound
\be
{|\de \tau|\over \tau_0}={2\pi\ga^2\over 15\tau_0}\int d\rb\ k\rb^2\leq{4\pi\over 45} \sqrt{6\pi}(\bM L)^3\approx 1.2 (\bM L)^3
\, , \ee
where we have used (\ref{tau0}) and~(\ref{gamma}).

%%%%%%%%%%%%%%%%%%%%%%%%%%%%%%%%%%%
\setcounter{equation}{0}
\section{Local Redshift Corrections} \label{App-redshift}
In 
this appendix  we try and obtain the redshift $z(r)$ corresponding to a 
source located at $r$ inside the patch and the observer  at the center. The differential equation governing this 
relation is given by
\be
{dz\over dr}=-{(1+z)\dot{R}'\over \sqrt{1+2E}}
\, . \ee
To estimate corrections inside the patch, which go $\sim \rb$, we can ignore $E(r)\sim \rb^2$. We then have
\begin{eqnarray}
{dz\over 1+z}=-\dot{R}' dr =-{2\pi\over 9}\ga^2\LT 
\tau^{-1}+2R_2\ga^2 \tau (kr)'\RT d\rb \nonumber \\
\approx - {2\pi\over 9}\ga^2\LT \tau_F^{-1}-{\pi\over 9}R_2\ga^4\LF{\tau_0\over \tau_F}\RF^2\rb k+2R_2\ga^2 \tau(k\rb)'\RT d\rb \, , \nonumber 
\end{eqnarray}
where we have used (\ref{tau-local}). The first term can be integrated 
in a straightforward manner to yield the FLRW result. The second term 
clearly is suppressed with respect to the third term by an extra factor of $\rb$ and therefore can be ignored. For the third term one can use the 
 trick of first integrating with respect to $r$ assuming that $\tau$ 
is a constant and then putting back the $r$-dependence of $\tau$ ({\it i.e.} we keep only leading powers of $r$).  After all 
these integrations we find
\be
\ln(1+z)-\ln C=-2\ln \tau_F(r)-{4R_2\pi\over 9}\ga^4 \tau_F k\rb
\, , \ee  
where $C$ is an integration constant which can be fixed  by demanding that at $r=0$, $z=0$. This gives us 
\be
\ln C=2\ln \tau_0
\, , \ee
so that we have
\be
1+z(r)\approx \LF{\tau_0\over \tau_F}\RF^2\exp\LT-{4R_2\pi\over 9}\ga^4 \tau k\rb \RT
\label{z(r)}
\, . \ee

%%%%%%%%%%%%%%%%%%%%%%%%%%%%%%%%%
\setcounter{equation}{0}
\section{From Newtonian to Synchronous Gauge} \label{gauges}
Our aim in this section is to try to cast the metric (\ref{eq:14}) in the small-$u$ and small-$E$ limit (or equivalently keeping upto linear terms in $k$)
\be
ds^2={ 9\tau^4\over \bM^2}\left\{-d\tau^2+\left[1+2R_2\ga^2\tau^2\left(\tir  k(\tir)\right)'-20 R_2\ga^2\tir^2 k(\tir)\right]d\tir^2+\left[1+2R_2\ga^2\tau^2  k(\tir)\right]d\Om^2\right\}
\label{LTBmetric}
\, , \ee
in a perturbative form:
\be
ds^2=a^2(\tau)[-(1+2\Phi)d\tau^2+(1-2\Phi)dx^2]
\, , 
\label{A-newtonian}
\ee
where $\Phi\ll 1$. We will see that this is possible by a suitable gauge transformation. 

In general, any metric containing only scalar fluctuations can be written as
\be
ds^2=a^2(\tau)\{-(1+2\psi)d\tau^2+2\p_i\omega d\tau dx^i+[(1-2\phi)\de_{ij}+D_{ij}\chi]dx^idx^j\}
\, , 
\label{A-metric}
\ee
where we have defined the operator
\be
D_{ij}\equiv \p_i\p_j-\3\de_{ij}\n^2
\, . \ee
The linear gauge transformations which leave the form of the metric and the scale factor invariant is given by
\begin{eqnarray}
\ti{\psi}&=&\psi-\ze_{\tau}-{a_{\tau}\over a}\ze
\\
\ti{\omega}&=&\omega+\ze+\bb_{\tau}
\\
\ti{\phi}&=&\phi-\3\n^2\bb +{a_{\tau}\over a}\ze
\\
\ti{\chi}&=&\chi+2\bb
\, , 
\label{A-gauge}
\end{eqnarray}
where the tilded variables are the new coordinate system, $\beta$ and $\zeta$ are two scalar parameters that define the gauge transformation, and the subscript $\tau$ indicates partial differentiation with respect to the conformal time $\tau$.

The strategy that we will adopt is to find appropriate gauge transformations which take us from the Newtonian gauge (\ref{A-newtonian}) to the synchronous (LTB) gauge. First of all, in order to deal with the LTB metric we have to work in the  radial coordinate system. To convert the different terms in (\ref{A-metric}) in the radial system we have to use the identities
\be
\p_i r={x_i\over r}
\, , \ee
and the differential relations
\be
x_idx^i=r dr \, , \qquad \mx{ and }\qquad  dx^idx_i=dr^2+r^2d\Om^2
\, , \ee

Then we find that when $\omega\ ,\phi$ are only functions of the radial coordinate, the terms involving them in the metric (\ref{A-metric}) are given by
\be
2\p_i\omega(r) dx^i=2\omega'(r){x_i\over r}dx^i=2\omega'(r)dr
\, , \ee
and
\be
(1-2\phi)\de_{ij}dx^idx^j=(1-2\phi)(dr^2+r^2d\Om^2) \, ,
\ee
while the term involving $\psi$ remains the same.

To compute the last term in (\ref{A-metric}) involving $\chi$ we need to evaluate the operator $D_{ij}$. Straight forwardly we find
\be
\p_i\p_j \chi(r)=\LT{x_ix_j\over r^2}\LF \chi''-{\chi'\over r}\RF+\de_{ij}{\chi'\over r}\RT
\, , \ee
and 
\be
\n^2\chi(r)=\chi''+{2\chi'\over r}
\, , \ee
so that 
\be
D_{ij}\chi dx^idx^j=\LF{x_ix_j\over r^2}-\3\de_{ij}\RF\LF \chi''-{\chi'\over r}\RF dx^idx^j={\cE}\LF {2\over 3}dr^2-\3 r^2d\Om^2\RF
\, , \ee
where we have defined
\be
\cE\equiv \chi''-{\chi'\over r}
\, . \ee
Thus putting everything together we have
\be
ds^2=a^2(\tau)\left[-(1+2\psi)d\tau^2+2\omega'(r)drd\tau +\LF1+2\phi+{2\over 3}\cE\RF dr^2+\LF 1+2\phi-\3{\cal E}\RF r^2d\Om^2\RT
\, 
\label{metric-radial}
. \ee

Now, for the Newtonian gauge, we have
\be
\psi=\phi=\Phi\mx{ and } \chi=\omega=0
\, . \ee
In order to go to the synchronous gauge we demand $\ti{\psi}$ and $\ti{\omega}$ in~(\ref{A-gauge}) to vanish, which  gives us
\be
\ze_{\tau}+{a_{\tau}\over a}\ze-\Phi=0\qquad \Ra \qquad \ze={\Phi(r)\tau\over 3}
\, , \ee
(here we ignore a possible integration constant), and
\be
\ze+\bb_{\tau}=0 \qquad \Rightarrow \qquad \bb=-{\Phi\tau^2\over 6}+\bb_0(r)
\, , \ee 
where the last term arises as an integration constant.
Under these gauge transformation the non-zero potentials become
\be
\ti{\chi}=0+2\bb=-{\Phi\tau^2\over 3}+2\bb_0(r)
\, \,  \Ra \, \, 
\ti{\cE}=-{\tau^2\over3}\LF\Phi''-{\Phi'\over r}\RF+2\LF\bb_0''-{\bb_0'\over r}\RF
\, , \ee
and
\be
\ti{\phi}=\Phi-\3\n^2\bb+{2\over \tau}\ze={5\over 3}\Phi+{1\over 18}\Phi''\tau^2+{1\over 9}{\Phi'\tau^2\over r}-{2\over 3}{\bb_0'\over r}-{1\over 3}\bb_0''
\label{ti-phi}
\, . \ee
We now have to check whether by appropriately choosing $\Phi(r)$ and $\bb_0(r)$ we can obtain $\ti{\cE}(r)$ and $\ti{\phi}$ corresponding to the LTB metric.  By comparing the LTB metric (\ref{LTBmetric}) with the metric in the radial coordinate system (\ref{metric-radial}) and using (\ref{ti-phi}) we find that this amounts to finding a $\{ \Phi(r),\bb_0(r)\}$ such that the following two equations are satisfied:
\be
-2\phi+{2\over 3}\cE=-{10\over 3}\Phi+2\bb_0''-\3\tau^2\Phi''=2R_2\ga^2\tau^2\left(\tir  k(\tir)\right)'-20 R_2\ga^2\tir^2 k(\tir)
\label{perteqn1}
\, , \ee
and
\be
-2\phi-\3{\cal E}=-{10\over 3}\Phi+2{\bb_0'\over \tir}-\3{\tau^2\Phi'\over \tir}=2R_2\ga^2\tau^2  k(\tir)
\, . \ee

The latter equation implies
\be
-{10\over 3}\Phi+2{\bb_0'\over \tir}=0\qquad \Ra \qquad \bb_0'={5\over 3}\Phi \tir
\, , \ee
and
\be
\Phi'=-6R_2\ga^2 \tir k\qquad \Ra \qquad \Phi=-6R_2\ga^2\int d\tir\ (\tir k )
\, . \ee
One can now easily check that the same solution set  $\{\Phi(\tir),\bb_0(\tir)\}$ also satisfies (\ref{perteqn1}). 
%%%%%%%%%%%%%%%%%%%%%%%%%%%%%%%%%%%
\setcounter{equation}{0}
\section{Estimating Corrections to $\tau$ in the General Case}\label{A-largeu}
The purpose in this section is to try and generalize the analysis done in the small curvature regime so that we can obtain an estimate for the correction for general density profiles. The general expressions for $R$ and $R'$ can be written as a power series in $u_0$, or $k(r)$:   
$$
R={\pi\over 3}r\ga^2\tau^2\LF1+\sum_1^{\infty}R_{2n} \ga^{2n}\tau^{2n} k^n\RF
$$
and
\be
R'={\pi\over 3}\ga^2\tau^2\LT1+\sum_1^{\infty}R_{2n} \ga^{2n}\tau^{2n} (rk^n)'\RT
\label{Rp-large}
\ee

The evolution equation for $\tau$ is then given by (\ref{t-radial}) and (\ref{Rp-large})
\be
{d\tau\over d\rb}={{\pi\over 9}\ga^2\LT1+\sum_1^{\infty}R_{2n} \ga^{2n}\tau^{2n} (\rb k^n)'\RT\over \sqrt{1+2k\rb^2}}\approx {\pi\over 9}\ga^2\LT1+\sum_1^{\infty}R_{2n} \ga^{2n}\tau^{2n} (rk^n)'+\sum_1^{\infty}S_nk^n\rb^{2n}\RT
\label{tau-series}
\, , \ee
where the coefficients $S_n$ are defined by the expansion
\be
(1+x)^{-1/2}\equiv 1+\sum_1^{\infty}S_nx^n
\, . \ee
From (\ref{tau-series}) we find the iterative expression
\be
\tau=\tau_F+\de\tau=\tau_F+{\pi \ga^2 \over 9}\LT\sum_1^{\infty} R_{2n}\ga^{2n}\int d\rb\ \tau^{2n}(\rb k^n)'+\sum_1^{\infty}S_n\int d\rb\  k^n\rb^{2n}\RT
\, . \ee

Now except for the $n=1$  term, all other ones in the $\{S_n\}$ series in (\ref{tau-series}) are  suppressed by a factor $\rb^3$ or more and therefore we can ignore them;  as we have explained before $\rb\ll 1$ and higher powers of $\rb$ will be suppressed. Further, the $n=1$ term has already been discussed in the small-$u$ case (finding that it goes like $(L/R_H)^3$), so we do not need to keep track of it. Thus we are left evaluating the $\{R_n\}$ series in (\ref{tau-series}). To get the leading non-zero correction, it is sufficient to keep  only upto $\cO(\rb^2)$  terms in the expansion of $\tau^{2n}$ that needs to be substituted in (\ref{tau-series}). This is straightforwardly obtained from the zeroth order (FLRW) result:
\be
\tau\approx\tau_0\LT1+{\pi \ga^2 r\bM\over 9}\RT\, \, \Ra \, \, \tau^{2n}\approx\tau_0^{2n}\LT1-{2n\pi \ga^2 r\bM\over 9\tau_0}+n(2n-1)\LF{\pi \ga^2 r\bM\over 9\tau_0}\RF^2\RT
\, , 
\ee 
where we have neglected once again higher powers of $r$.
So, $\de\tau$ now reads
\be
\de \tau\approx {\pi \ga^2 \over 9}\sum_1^{\infty}R_{2n}\ga^{2n}\tau_0^{2n}\int d\rb \LT1-{2n\pi \ga^2 \rb\over 9\tau_0}+n(2n-1)\LF{\pi \ga^2 \rb\over 9\tau_0}\RF^2\RT(rk^n)'
\, . 
\label{A-deltau}
\ee

We are only interested in estimating the correction when the photon goes from boundary to boundary. As in the small-$u$ case, the first term, being an integral of a total derivative,  vanishes when the integral is evaluated from a boundary to another boundary.  Next let us try to evaluate the contribution from the second term. It is of the form
$$\int r\bM(rk^n)'dr=\bM\LT r^2k^n-\int dr\ rk^n\RT$$
The first term again vanishes at the boundary or at the center. Moreover, since the integrand in  the second term is odd,  the integral also vanishes when evaluated from boundary to boundary.

Thus we can only hope to obtain a non-zero contribution from the third  term in (\ref{A-deltau}). Let us therefore look at the integrals
\be
\int \rb^2(\rb k^n)'d\rb=\LT \rb^3k^n-2\int d\rb\ \rb^2k^n\RT
\, . \ee
Now, the integrand in the r.h.s. is even and therefore it does not vanish. This is the first non-zero correction coming from the inhomogeneities and  gives us
\be
{\de \tau\over\tau_0}\approx -\LF{\pi \ga^2 \over 9\tau_0}\RF^3\sum_1^{\infty} R_{2n}\ga^{2n}\tau_0^{2n}2n(2n-1)\int_{-L}^L d\rb\ \rb^2k^n
\, . \ee
One can try to find an upper bound for the above expression as follows:
\be
\int d\rb\ \rb^2k^n\leq {2\over 3} \kx^n (\bM L)^3
\, . \ee
Thus we have 
\be
{|\de \tau|\over \tau_0}\leq \LF{\bM L\pi \ga^2 \over 9\tau_0}\RF^3\sum_1^{\infty} 2n(2n-1) R_{2n}\tau_0^{2n}\ga^{2n}\kx^n 
\, . \ee

In order to estimate the sum, first we observe that the sum can be written in terms of a function $f(u_0)$:
\be
f(u_0)=\sum_1^{\infty} R_{2n}u_0^{2n}\, \, \Ra \, \,  u_0^2{d^2f\over du_0^2}=\sum_1^{\infty} 2n(2n-1) R_{2n}u_0^{2n}=\sum_1^{\infty} 2n(2n-1) R_{2n}\tau_0^{2n}\ga^{2n}k^n 
\, . \ee
Therefore we have
\be
{|\de \tau|\over \tau_0}\leq (\bM L)^3\LF{\pi \ga^2 \over 9\tau_0}\RF^3  \LF u_0^2{d^2f\over du_0^2}\right)_{\mt{maximum}} \, . \ee

%%%%%%%%%%%%%%%%%%%%%%%%%%%%


\begin{thebibliography}{99}







%\cite{Biswas:2006ub}
\bibitem{BMN}
  T.~Biswas, R.~Mansouri and A.~Notari,
  %``Nonlinear Structure Formation and Apparent Acceleration: an
  %Investigation,''
  arXiv:astro-ph/0606703.
  %%CITATION = ASTRO-PH 0606703;%%



%\cite{Albrecht:2006um}
\bibitem{taskforce}
  A.~Albrecht {\it et al.},
  %``Report of the Dark Energy Task Force,''
  arXiv:astro-ph/0609591.
  %%CITATION = ASTRO-PH 0609591;%%

\bibitem{buchert}
  T.~Buchert and J.~Ehlers,
  %``Averaging Inhomogeneous Newtonian Cosmologies,''
  Astron.\ Astrophys.\  {\bf 320}, 1 (1997)
  [arXiv:astro-ph/9510056]; T.~Buchert,
  %``On average properties of inhomogeneous fluids in general relativity.  I:
  %Dust cosmologies,''
  Gen.\ Rel.\ Grav.\  {\bf 32}, 105 (2000)
  [arXiv:gr-qc/9906015].
  %%CITATION = GR-QC 9906015;%%


\bibitem{rasanen}
  S.~Rasanen,
  %``Dark energy from backreaction,''
  JCAP {\bf 0402}, 003 (2004)
  [arXiv:astro-ph/0311257].
  %%CITATION = ASTRO-PH 0311257;%%

\bibitem{KMNR}
  E.~W.~Kolb, S.~Matarrese, A.~Notari and A.~Riotto,
  %``The effect of inhomogeneities on the expansion rate of the universe,''
  Phys.\ Rev.\ D {\bf 71}, 023524 (2005)
  [arXiv:hep-ph/0409038].
  %%CITATION = HEP-PH 0409038;%%

\bibitem{notari}
  A.~Notari,
  %``Late time failure of Friedmann equation,''
  Mod.\ Phys.\ Lett.\  A {\bf 21}, 2997 (2006)
  [arXiv:astro-ph/0503715].
  %%CITATION = MPLAE,A21,2997;%%

\bibitem{greci}
  N.~Brouzakis, N.~Tetradis and E.~Tzavara,
  %``The Effect of Large-Scale Inhomogeneities on the Luminosity Distance,''
  arXiv:astro-ph/0612179.
  %%CITATION = ASTRO-PH/0612179;%%

\bibitem{subir}
  A.~Blanchard, M.~Douspis, M.~Rowan-Robinson and S.~Sarkar,
  %``An alternative to the cosmological 'concordance model',''
  Astron.\ Astrophys.\  {\bf 412}, 35 (2003)
  [arXiv:astro-ph/0304237].
  %%CITATION = ASTRO-PH 0304237;%%


%\cite{Celerier:1999hp}
\bibitem{celerier}
  M.~N.~Celerier,
  %``Do we really see a Cosmological Constant in the Supernovae data ?,''
  Astron.\ Astrophys.\  {\bf 353}, 63 (2000)
  [arXiv:astro-ph/9907206].
  %%CITATION = AAEJA,353,63;%%

\bibitem{Tomita00} K. Tomita, Astrophys. J., 529, 38, 2000; astro-ph/0005031.
\bibitem{Tomita01} K. Tomita, Prog. Theor. Phys. 106, No. 5, 2001.



\bibitem{wiltshire} D. L. Wiltshire, gr-qc/0503099.

\bibitem{moffat} %\cite{Moffat:2005yx}
  J.~W.~Moffat,
  %``Cosmic Microwave Background, Accelerating Universe and Inhomogeneous
  %Cosmology,''
  JCAP {\bf 0510}, 012 (2005)
  %%CITATION = JCAPA,0510,012;%%
astro-ph/0502110 %\cite{Moffat:2005ii}
  \, and JCAP {\bf 0605}, 001 (2006)
  %%CITATION = JCAPA,0605,001;%%

\bibitem{alnes} %\cite{Alnes:2005rw}
  H.~Alnes, M.~Amarzguioui and O.~Gron,
  %``An inhomogeneous alternative to dark energy?,''
  Phys.\ Rev.\  D {\bf 73}, 083519 (2006)
  [arXiv:astro-ph/0512006].
  %%CITATION = PHRVA,D73,083519;%%


%\cite{Mansouri:2005rf}
\bibitem{reza}
  R.~Mansouri,
  %``Structured FRW universe leads to acceleration: A non-perturbative
  %approach,''
  arXiv:astro-ph/0512605.
  %%CITATION = ASTRO-PH/0512605;%%


\bibitem{INOUESILK}
  K.~T.~Inoue and J.~Silk,
  %``Local Voids as the Origin of Large-angle Cosmic Microwave Background
  %Anomalies: The Effect of a Cosmological Constant,''
  arXiv:astro-ph/0612347;
  K.~T.~Inoue and J.~Silk,
  %``Local Voids as the Origin of Large-angle Cosmic Microwave Background
  %Anomalies,''
  arXiv:astro-ph/0602478.
  %%CITATION = ASTRO-PH 0602478;%%

\bibitem{hole}
  W.~J.~Frith, G.~S.~Busswell, R.~Fong, N.~Metcalfe and T.~Shanks,
  %``The Local Hole in the Galaxy Distribution: Evidence from 2MASS,''
  Mon.\ Not.\ Roy.\ Astron.\ Soc.\  {\bf 345}, 1049 (2003)
  [arXiv:astro-ph/0302331].
  %%CITATION = ASTRO-PH 0302331;%%


\bibitem{ReesSciama}
	M.~J.~Rees and D.~W.~Sciama,
	Nature {\bf 517}, 611 (1968).


\bibitem{MGSS}
  E.~Martinez-Gonzalez, J.~L.~Sanz and J.~Silk,
  %``Minimal anisotropies in the cosmic microwave background,''
  Phys.\ Rev.\  D {\bf 46}, 4193 (1992)%\cite{Martinez-Gonzalez:1994ca}
  %``Imprints Of Galaxy Clustering Evolution On Delta T/T,''
  \, and arXiv:astro-ph/9406001.
  %%CITATION = ASTRO-PH/9406001;%%
  %%CITATION = PHRVA,D46,4193;%%



\bibitem{voids}
  H.~J.~Rood,
  %``Voids,''
  Ann.\ Rev.\ Astron.\ Astrophys.\  {\bf 26}, 245 (1988)
  %%CITATION = ARAAA,26,245;%%
;
S.~G.~Patiri, J.~Betancort-Rijo, F.~Prada, A.~Klypin and S.~Gottlober,
  %``Statistics of Voids in the 2dF Galaxy Redshift Survey,''
  Mon.\ Not.\ Roy.\ Astron.\ Soc.\  {\bf 369}, 335 (2006)
  [arXiv:astro-ph/0506668];
  F.~Hoyle and M.~S.~Vogeley,
  %``Voids in the 2dF Galaxy Redshift Survey,''
  Astrophys.\ J.\  {\bf 607}, 751 (2004)
  [arXiv:astro-ph/0312533].
  %%CITATION = ASJOA,607,751;%%






\bibitem{LTB} G. Lema\^itre, Ann. soc. Sci. Bruxelles Ser.1, A53, 51, 1933; 
              R. C. Tolman,Proc. Nat1. Acad. Sci. U.S.A. 20,410, 1934; 
              H. Bondi, Mon. Not. R. Astron. Soc., 107, 343, 1947).

\bibitem{reza2}
  S.~Khakshournia and R.~Mansouri,
  %``Formation of cosmological mass condensation within a FRW universe: exact
  %general relativistic solutions,''
  Phys.\ Rev.\  D {\bf 65}, 027302 (2002)
  [arXiv:gr-qc/0307023].
  %%CITATION = PHRVA,D65,027302;%


\bibitem{Flanagan}
  R.~A.~Vanderveld, E.~E.~Flanagan and I.~Wasserman,
  %``Mimicking Dark Energy with Lema\^itre-Tolman-Bondi Models: Weak Central
  %Singularities and Critical Points,''
  arXiv:astro-ph/0602476.
  %%CITATION = ASTRO-PH 0602476;%%


\bibitem{zeldovich}
  Y.~B.~Zeldovich,
  %``Gravitational Instability: An Approximate Theory For Large Density
  %Perturbations,''
  Astron.\ Astrophys.\  {\bf 5} (1970) 84.
  %%CITATION = AAEJA,5,84;%%


\bibitem{future}
T.~Biswas and A.~Notari, work in progress.


%\cite{Mukhanov:1990me}
\bibitem{MFB}
  V.~F.~Mukhanov, H.~A.~Feldman and R.~H.~Brandenberger,
  %``Theory of cosmological perturbations. Part 1. Classical perturbations. Part %2. Quantum theory of perturbations. Part 3. Extensions,''
  Phys.\ Rept.\  {\bf 215}, 203 (1992).
  %%CITATION = PRPLC,215,203;%%


\bibitem{vishniac}
K.~L.~Thompson and   E.~T.~Vishniac, Ap.J {\bf 313}, 517 (1987).

\bibitem{panek}
	M.~Panek, Ap J., {\bf 388}, 225-233 (1992).

%\cite{Fullana:1996vj}
\bibitem{fullana}
  M.~J.~Fullana, J.~V.~Arnau and D.~Saez,
  %``On the microwave background anisotropy produced by big voids in open
  %universes,''
  arXiv:astro-ph/9601154.
  %%CITATION = ASTRO-PH/9601154;%%

\bibitem{arnau}	
 J.~V.~Arnau ; M.~J.~Fullana; L.~Monreal; D.~Saez,	
	
	Ap.J. {\bf 402} 359-368 (1993).


%\cite{Mollerach:1997up}
\bibitem{MM}
  S.~Mollerach and S.~Matarrese,
  %``Cosmic microwave background anisotropies from second order  gravitational
  %perturbations,''
  Phys.\ Rev.\ D {\bf 56}, 4494 (1997)
  [arXiv:astro-ph/9702234].
  %%CITATION = ASTRO-PH 9702234;%%























\end{thebibliography}
\end{document}